\documentclass[aps,pra,twocolumn,floatfix,showpacs,preprintnumbers,amsmath,amssymb,10pt]{revtex4-1}
\usepackage[dvips]{graphics}

\DeclareMathAlphabet{\mathsfit}{\encodingdefault}{\sfdefault}{m}{sl}
\SetMathAlphabet{\mathsfit}{bold}{\encodingdefault}{\sfdefault}{bx}{sl}

\usepackage{mathptmx}
\usepackage{psfrag}
\usepackage{graphicx}
\usepackage{bm}
\usepackage{amsmath}
\usepackage{trfsigns}
\usepackage{amssymb}
\usepackage[usenames,dvipsnames]{color}
\usepackage{dashrule}
\usepackage{textgreek}
\usepackage[english]{babel}
\usepackage{contour}
\usepackage{upgreek,bm}
\usepackage{color}
\usepackage{dsfont}
\definecolor{dred}{rgb}{.6,.0,0.}
\definecolor{dblue}{rgb}{.0,.0,0.6}

\newcommand{\bra}[1]{\ensuremath{\left\langle#1\right|}}
\newcommand{\ket}[1]{\ensuremath{\left|#1\right\rangle}}
\renewcommand{\vec}[1]{\mathbf{#1}}

\newcommand{\tens}[1]{\mbox{\textsf{\textbf{#1}}}}

\newcommand{\Greektens}[1]{\contour[3]{black}{#1}}

\newcommand{\sprod}{\!\cdot\!}

\newcommand{\vprod}{\!\times\!}

\newcommand{\dif}{\mathrm{d}}
\newcommand{\mi}{\textrm{i}} 
\newcommand{\me}{\mathrm{e}}

\begin{document}

\title{Casimir--Polder Potential of a Driven Atom}

\author{Sebastian Fuchs$^1$}
\author{Robert Bennett$^1$}
\author{Stefan Yoshi Buhmann$^{1,2}$}
\affiliation{$^1$ Physikalisches Institut, Albert-Ludwigs-Universit\"at Freiburg, Hermann-Herder-Stra{\ss}e 3, 79104 Freiburg, Germany\\
$^2$ Freiburg Institute for Advanced Studies, Albert-Ludwigs-Universit\"at Freiburg, Albertstra{\ss}e 19, 79104 Freiburg, Germany}

\date{\today}

\begin{abstract}
We investigate theoretically the Casimir--Polder potential of an atom which is driven by a laser field close to a surface. This problem is addressed in the framework of macroscopic quantum electrodynamics using the Green's tensor formalism and we distinguish between two different approaches, a perturbative ansatz and a method based on Bloch equations. We apply our results to a concrete example, namely an atom close to a perfectly conducting mirror, and create a scenario where the tunable Casimir--Polder potential becomes similar to the respective potential of an undriven atom due to fluctuating field modes. Whereas the perturbative approach is restricted to large detunings, the ansatz based on Bloch equations is exact and yields an expression for the potential which does not exceed $1/2$ of the undriven Casimir--Polder potential. 
\end{abstract}

\maketitle

\section{Introduction}
Casimir--Polder forces \cite{Casimir_Polder:1948} are weak electromagnetic forces between an atom and a surface caused by spontaneously arising noise currents both in the atom and the surface. These noise currents are the source of quantized electromagnetic fields, which are described by the theory of macroscopic Quantum Electrodynamics (QED) \cite{Buhmann_Book_1, Buhmann_Book_2}. This extension of vacuum QED incorporates the presence of macroscopically modeled matter in its field operators. Electric and magnetic fields are given by these field operators and the classical dyadic Green's tensor which contains the physical and geometrical information regarding the surface. The Green's tensor is the propagator of the electromagnetic field and mathematically, it is the formal solution of the Helmholtz equation \cite{Dung:1998, Buhmann:2004, Buhmann:2007, Scheel:2008}. The surface's presence causes a frequency shift \cite{Wylie:1984, Wylie:1985} in the atomic transition frequency which is the reason for a usually attractive force of the atom towards the surface, the Casimir--Polder force, cf. e.g Ref.~\cite{Buhmann:2004}.\\
Casimir--Polder shifts and potentials have been studied extensively for a huge variety of different physical setups and configurations. Different materials, such as metals \cite{Wylie:1984}, graphene \cite{Eberlein:2012} and metamaterials~\cite{Xu:2014, Henkel:2005} have been studied, whereas e.g.~Casimir--Polder potentials for nonreciprocal materials \cite{Fuchs:2016} require an extension of the theory \cite{Buhmann_nonreciprocal:2012}. Moreover one can study atoms in the ground state or the excited state, in an environment at $T=0$ \cite{Casimir_Polder:1948} or at a temperature different from zero \cite{McLachlan:1963}. Additionally there is a static way of calculating potentials using perturbation theory \cite{McLachlan:1963} and a dynamical way by solving the internal atomic dynamics \cite{Milonni:1982}.\\
Experimentally, there are several approaches to measuring Casimir--Polder forces and verifying the developed theories. One of the first approaches \cite{Sukenik:1993} is based on a measurement of the deflection of atoms passing through a parallel-plate cavity as a function of plate separation. The Casimir--Polder force can inferred by measuring the angle of deflection. If the incoming atoms are very slow and are reflected by the medium the scattering process has to be described quantum mechanically \cite{Friedrich:2002}. A respective experiment is presented in Ref.~\cite{Shimizu:2001}. Another method is the study of mechanical motions of a Bose-Einstein condensate (BEC) under the influence of a surface potential \cite{Harber:2005}. The Rb BEC is trapped magnetically and the perturbation of the the center-of-mass oscillations due to the surface's presence are detected. Similar to the method mentioned above the temperature-dependence of Casimir--Polder Forces was investigated \cite{Obrecht:2007}. In this experiment a Bose--Einstein condensate of Rb atoms was brought close to a dielectric substrate and the collective oscillation frequency of its mechanical dipole was measured. Higher temperatures are generated by heating the substrate with a laser. At close distances the effect of the Casimir--Polder force on the trap potential is significant.\\
These Casimir--Polder forces are present for single atoms that are trapped next to a surface. Ref.~\cite{Thompson:2013} presents an experiment where a single Rb atom that is trapped by a tightly focused optical tweezer beam \cite{Schlosser:2001} couples to a solid-state device, namely a nanoscale photonic crystal cavity. The trap is essentially a standing wave formed by the laser beam and its reflected beam with minima of potential energy at the intensity maxima. At a low temperature a single atom is loaded into the first minimum of potential energy by scanning the optical tweezer over the surface \cite{Thompson:2013,Thompson:2013_2}. The atom's position can be controlled precisely, until the atom comes too close to the surface where the attractive Casimir--Polder potential dominates over the trap potential. Significant effects of the Casimir--Polder force on the trapping lifetime of atoms was already predicted for magnetically trapped atoms close to a surface \cite{Fermani:2007}.\\
In this context we want to mention experiments \cite{Balykin:1988, Oberst:2003} using atomic beams and a laser to reflect the atomic beam next a dielectric. The laser field is internally reflected at the dielectric's surface producing a thin wave along the surface, which decays exponentially in the normal direction. An incoming atom feels a gradient force in this surface wave expelling the atom out of the field with a detuning. This reflection process is state-selective \cite{Balykin:1988}. Such a setup can also be used to measure the Casimir--Polder force between ground-state atoms and a mirror~\cite{Landragin:1996}. Laser-cooled atoms with a specific kinetic energy are brought close to the mirror with evanescent wave. The atoms are reflected from an evanescent wave atomic mirror if their kinetic energy is higher than the potential barrier. By measuring the kinetic energy of the atoms, the intensity and the detuning of the evanescent wave, the CP force can be extracted.\\
In this work, we want to investigate the Casimir--Polder potential for a laser-driven atom and study this problem in off-resonant and resonant regimes. The solution is described in the framework of macroscopic QED using Green's tensors. The off-resonant regime has previously been studied in Ref.~\cite{Perreault:2008} for a perfect conductor. However, as we will show in the following, the obtained results only hold in the nonretarded regime. A similar calculation based on the optical Bloch equations, as in the resonant regime, is carried out in Ref.~\cite{Huang:1984}. Ref.~\cite{Obrecht:2007_2} reports of an experiment with resonantly driven atoms that are already adsorbed on a surface. It is possible to measure the electric fields generated by these atoms.\\
This paper is organized as follows. Sec.~\ref{sec:Macroscopic Quantum Electrodynamics} gives an overview over the interaction Hamiltonian and describes the decomposition of the electric field into a free and an induced contribution containing the Green's tensor. The internal atomic dynamics is outlined in Sec.~\ref{sec:Internal Atomic Dynamics}, where the surface-induced frequency shift and decay rate are introduced. We distinguish here between a perturbative approach in Sec.~\ref{sec:Perturbative Approach} and \ref{sec:Components of the Electric Potential} and an ansatz based on Bloch equations \ref{sec:Bloch Equation}. We derive dipole moments in the time domain, the free laser force and the Casimir--Polder potential for both methods. Sec.~\ref{sec:Atom Near a Plane Surface} studies the example of a two-level atom in front of a perfectly conducting mirror and gives a comparison of both methods with the undriven standard Casimir--Polder potential.

\section{The Electric Field in Macroscopic Quantum Electrodynamics}
\label{sec:Macroscopic Quantum Electrodynamics}
We compute the Casimir--Polder potential for an atom that is driven by a laser field in the presence of a surface. The theory of macroscopic QED is an extension of vacuum QED that incorporates the surface in its field operators. This system is governed by a Hamiltonian $\hat H$ consisting of an atomic part $\hat{H}_{\textrm{A}}$, a field part $\hat{H}_{\textrm{F}}$ containing surface effects and the interaction part between the atom and the modified field $\hat{H}_{\textrm{AF}}$. The field part of the Hamiltonian $H{}_{\textrm{F}}$
\begin{equation}
\hat{H}{}_{\textrm{F}} = \sum\limits_{\lambda = \textrm{e,m}}{\int{\dif^3 r \int\limits^{\infty}_0{\dif \omega \hbar \omega \hat{\vec{f}}{}^{\dagger}{}_{\lambda} \left( \vec{r}, \omega \right) \sprod \hat{\vec{f}}{}_{\lambda} \left( \vec{r}, \omega \right) }}}
\label{eq:Field Hamiltonian}
\end{equation}
sums over both electric and magnetic fundamental excitations $\textrm{e}, \textrm{m}$ and integrates the matter-modified creation and annihilation operators $\hat{\vec{f}}{}^{\dagger}{}_{\lambda} \left( \vec{r}, \omega \right)$, $\hat{\vec{f}}{}_{\lambda} \left( \vec{r}, \omega \right)$ of the body-field system over the entire space in position and frequency. The electric and magnetic excitations establish the spontaneously arising noise polarization $\hat{\vec{P}}{}_{\textrm{N}}$ and noise magnetization $\hat{\vec{M}}{}_{\textrm{N}}$, respectively, which together form the noise current $\hat{\vec{j}}{}_{\textrm{N}} = -\mi \omega \hat{\vec{P}}{}_{\textrm{N}} + \overrightarrow{\nabla} \vprod \hat{\vec{M}}{}_{\textrm{N}}$. These fluctuating noise currents of the matter-field system are the origin of electric and magnetic fields in a variety of dispersion forces, such as the van der Waals force between the electronic shells of two atoms/molecules and the Casimir force between macroscopic objects. These dressed bosonic field operators follow the commutation relations
\begin{equation}
\left[ \hat{\vec{f}}{}_{\lambda} \left( \vec{r}{}, \omega \right), \hat{\vec{f}}_{\lambda'}^{\dagger}\left( \vec{r}{}', \omega' \right) \right] = \delta_{\lambda \lambda'} \textrm{\Greektens{$\delta$}} \left( \vec{r}- \vec{r}{}' \right) \delta \left( \omega - \omega' \right).
\end{equation}
Acting on the ground state $\ket{\{ 0 \}}$ the annihilation operator $\hat{\vec{f}}{}_{\lambda} \left( \vec{r}{}, \omega \right)$ gives $0$ for all values of $\lambda$, $\omega$ and $\vec{r}$. Higher field states are produced by acting $\hat{\vec{f}}_{\lambda}^{\dagger} \left( \vec{r}{}, \omega \right)$ on the ground state of the field $\ket{\{ 0 \}}$.\\
The expression for the electric field is defined by the field operators and the dyadic Green's function, named Green's tensor \cite{Buhmann_Book_2}
\begin{align}
\begin{array}{lll}
&\hat{\vec{E}} \left( \vec{r}, t \right) &= \int\limits^{\infty}_0{\dif \omega \left[ \hat{\vec{E}} \left( \vec{r}, \omega, t \right) + \hat{\vec{E}}{}^{\dagger} \left( \vec{r}, \omega, t \right) \right]}\\
& &= \int\limits^{\infty}_0{\dif \omega \sum\limits_{\lambda = \textrm{e,m}}{\int{\dif^3 \vec{r}{}' \tens{G}{}_{\lambda} \left( \vec{r}, \vec{r}{}', \omega \right) \sprod \hat{\vec{f}}{}_{\lambda} \left( \vec{r}{}', \omega \right)}}} + \textrm{h.c.}
\label{eq:Electric Field Greens Tensor}
\end{array}
\end{align}
The Green's tensor formally solves the Helmholtz equation for the electric field resulting from the Maxwell equations in vacuum. It can be considered as the field propagator between field points and source points and can also be decomposed into electric and magnetic contributions satisfying the integral relation
\begin{equation}
\sum\limits_{\lambda = \textrm{e,m}}{\int{\dif^3 s \tens{G}_{\lambda} \left( \vec{r}, \vec{s}, \omega \right) \sprod \tens{G}^{*\textrm{T}}_{\lambda} \left( \vec{r}{}', \vec{s}, \omega \right)}} = \frac{\hbar \mu_0}{\pi} \omega^2 \textrm{Im} \tens{G} \left( \vec{r}, \vec{r}{}', \omega \right).
\end{equation}
The atomic Hamiltonian
\begin{equation}
\hat{H}{}_{\textrm{A}} = \sum\limits_n{E_n \hat{A}{}_{nn}}
\label{eq:Atomic Hamiltonian}
\end{equation}
contains the atomic eigenenergies $E_n$ of level $n$ and the diagonal elements of the atomic flip operator $\hat{A}{}_{mn} = \ket{m} \bra{n}$.\\
The interaction Hamiltonian $\hat{H}{}_{\textrm{AF}}$ contains the electric dipole moment operator $\hat{\vec{d}}$ that can be represented in terms of the atomic flip operator
\begin{equation}
\hat{\vec{d}} = \sum\limits_{m,n}{\vec{d}_{mn} \hat{A}{}_{mn}}.
\label{eq:Dipole Moment Definition}
\end{equation}
Using the expression for the dipole moment, the atom-field interaction Hamiltonian reads
\begin{equation}
\hat{H}{}_{\textrm{AF}} = - \hat{\vec{d}} \sprod \hat{\vec{E}} \left( \vec{r}{}_{\textrm{A}} \right) = - \sum\limits_{m,n}{\hat{A}{}_{mn} \vec{d}{}_{mn} \sprod \hat{\vec{E}} \left( \vec{r}{}_{\textrm{A}} \right)},
\label{eq:Interaction Hamiltonian}
\end{equation}
where $\vec{r}_{\textrm{A}}$ is the atom's position. Inserting Eq.~\eqref{eq:Electric Field Greens Tensor} into the interaction Hamiltonian \eqref{eq:Interaction Hamiltonian} allows us to set up the Heisenberg equation of motion for the field operator by using the total Hamiltonian of the system \eqref{eq:Field Hamiltonian}, \eqref{eq:Atomic Hamiltonian} and \eqref{eq:Interaction Hamiltonian}, whose solution reads
\begin{multline}
\hat{\vec{f}}_{\lambda} \left( \vec{r}, \omega, t \right) = \me^{- \mi \omega \left( t-t_0 \right)} \hat{\vec{f}}_{\lambda} \left( \vec{r}, \omega \right)\\
+ \frac{\mi}{\hbar} \int\limits^t_{t_0}{\dif t' \me^{-\mi \omega \left( t-t' \right)} \tens{G}{}^{*\textrm{T}}_{\lambda} \left( \vec{r}{}_{\textrm{A}}, \vec{r}, \omega \right) \sprod \hat{\vec{d}} \left( t' \right)}.
\label{eq:Field Operator}
\end{multline}
The field operator $\hat{\vec{f}}_{\lambda} \left( \vec{r}, \omega, t \right)$ in the Heisenberg picture evaluated at time $t_0$ would reproduce the time-independent equivalent in the Schr\"{o}dinger picture.\\
The first part of the annihilation operator is the free contribution in absence of the atom. At the laser source $V_{\textrm{S}}$ it has a coherent-state contribution of the laser field and otherwise there are ubiquitous vacuum fluctuations. A corresponding field state reads
\begin{equation}
\ket{\psi}_{\textrm{F}} = \underset{\vec{r} \in V_{\textrm{S}}}{\ket{\left\{ \vec{f}_{\lambda} \left( \vec{r}, \omega \right) \right\}}} \otimes \underset{\vec{r} \notin V_{\textrm{S}}}{\ket{ \left\{ 0 \right\}}}.
\label{eq:Quantum State}
\end{equation}
If the annihilation operator $\hat{\vec{f}}_{\lambda} \left( \vec{r}, \omega, t \right)$ from Eq.~\eqref{eq:Field Operator} acts on the state \eqref{eq:Quantum State}, there are consequently two contributions
\begin{equation}
\hat{\vec{f}}_{\lambda} \left( \vec{r}, \omega \right) \ket{\psi}_{\textrm{F}} = \begin{cases} \vec{f}_{\lambda} \left( \vec{r}, \omega \right) \ket{\psi}_{\textrm{F}} & \textrm{if} \; \vec{r} \in V_{\textrm{S}}\\ 0 & \textrm{if} \; \vec{r} \notin V_{\textrm{S}}. \end{cases}
\label{eq:Electric Field Quantum State}
\end{equation}
Such a deconvolution of the field operators was done in Ref.~\cite{Dalibard:1982}. The result \eqref{eq:Field Operator} can be inserted into the equation for the electric field \eqref{eq:Electric Field Greens Tensor} yielding the final expression for the time-dependent electric field operator
\begin{multline}
\hat{\vec{E}} \left( \vec{r}, \omega, t \right) = \hat{\vec{E}}{}_{\textrm{free}} \left( \vec{r}, \omega, t \right) + \hat{\vec{E}}{}_{\textrm{ind}} \left( \vec{r}, \omega \right)\\
=\hat{\vec{E}} \left( \vec{r}, \omega \right) \me^{- \mi \omega \left( t - t_0 \right)}\\
+ \frac{\mi \mu_0}{\pi} \omega^2 \int\limits^t_{t_0}{\dif t' \me^{-\mi \omega \left( t-t' \right)} \mathrm{Im} \tens{G} \left( \vec{r}, \vec{r}{}_{\textrm{A}}, \omega \right) \sprod \hat{\vec{d}} \left( t' \right)}
\label{eq:Electric Field Free and Induced}
\end{multline}
with the free component
\begin{align}
\begin{array}{lll}
\hat{\vec{E}}_{\textrm{free}} \ket{\psi}_{\textrm{F}} &=& \int \limits_{V_{\textrm{S}}} \dif^3 r' \tens{G}_{\lambda} \left( \vec{r}, \vec{r} \; ', \omega \right) \sprod \vec{f}_{\lambda} \left( \vec{r}, \omega \right) \ket{\psi}_{\textrm{F}}\\
&\equiv& \vec{E} \left( \vec{r}, \omega \right) \ket{\psi}_{\textrm{F}}.
\end{array}
\end{align}
The classical electric driving field of the laser at the atom's position $\vec{E} \left( \vec{r}_{\textrm{A}}, t \right)$ can be written as Fourier-relations with time-independent and time-dependent frequency components $\vec{E} \left( \vec{r}_{\textrm{A}}, \omega, t \right)$ and $\vec{E} \left( \vec{r}_{\textrm{A}}, \omega \right)$, similar to Eq.~\eqref{eq:Electric Field Greens Tensor},
\begin{align}
\begin{array}{lll}
&\vec{E} \left( \vec{r}_{\textrm{A}}, t \right) &= \int\limits^{\infty}_0{\dif \omega \left[ \vec{E} \left( \vec{r}_{\textrm{A}}, \omega, t \right) + \vec{E}^* \left( \vec{r}_{\textrm{A}}, \omega, t \right) \right]}\\[4mm]
& &= \int\limits^{\infty}_0{\dif \omega \left[ \me^{-\mi \omega t} \vec{E} \left( \vec{r}_{\textrm{A}}, \omega \right) + \me^{\mi \omega t} \vec{E}^* \left( \vec{r}_{\textrm{A}}, \omega \right) \right]}\\[4mm]
& &= \vec{E} \left( \vec{r}_{\textrm{A}} \right) \cos \left( \omega_{\textrm{L}} t \right)
\label{eq:Electric Driving Field}
\end{array}
\end{align}
with the driving frequency of the laser $\omega_{\textrm{L}}$. The frequency components can then be identified as
\begin{align}
\begin{array}{lll}
&\vec{E} \left( \vec{r}_{\textrm{A}}, \omega \right) &= \frac{1}{2} \vec{E} \left( \vec{r}_{\textrm{A}} \right) \delta \left( \omega - \omega_{\textrm{L}} \right),\\[2mm]
&\vec{E}^* \left( \vec{r}_{\textrm{A}}, \omega \right) &= \frac{1}{2} \vec{E} \left( \vec{r}_{\textrm{A}} \right) \delta \left( \omega - \omega_{\textrm{L}} \right).
\label{eq:Frequency Components Driving Field}
\end{array}
\end{align}
The second part of Eq.~\eqref{eq:Electric Field Free and Induced} is the induced field stemming from the atom directly. This term is affected by the atom's position $\vec{r}_{\textrm{A}}$ and state at all times after the preparation into the initial state.\\
The induced part of the electric field $\hat{\vec{E}}_{\textrm{ind}} \left( \vec{r}, \omega \right)$ in Eq.~\eqref{eq:Electric Field Free and Induced} depends on the dipole moment of the atom $\hat{\vec{d}} \left( t \right)$. In Sec.~\ref{sec:Internal Atomic Dynamics}, where the internal atomic dynamics is investigated, the dipole moment is split into a free fluctuating part and an induced part as well. Following perturbation theory, the induced electric field $\hat{\vec{E}}_{\textrm{ind}} \left( \vec{r}, \omega \right)$ depend on the free dipole moment. Only higher terms would contain the induced contributions again. The procedure of decomposing the electric field and the dipole operator into free and induced parts related to the order of perturbation is taken from Refs.~\cite{Vasile:2008, Messina:2010, Haakh:2014}.\\
In Sec.~\ref{sec:Perturbative Approach} the induced dipole moment and the induced electric field are computed in a perturbative approach.

\section{Internal Atomic Dynamics}
\label{sec:Internal Atomic Dynamics}
After deriving an expression for the electric field consisting of the free part and the induced part in Sec.~\ref{sec:Macroscopic Quantum Electrodynamics}, one can compute the Heisenberg equation of motion for the atomic flip operator $\hat{A}{}_{mn} \left( t \right)$ in a similar way \cite{Buhmann_Book_2}
\begin{multline}
\dot{\hat{A}}{}_{mn} \left( t \right) = \mi \omega_{mn} \hat{A}{}_{mn} \left( t \right)\\
+ \frac{\mi}{\hbar} \sum\limits_k \int^{\infty}_0 \dif \omega \left\{ \left[ \hat{A}{}_{mk} \left( t \right) \vec{d}{}_{nk} - \hat{A}{}_{kn} \left( t \right) \vec{d}_{km} \right] \sprod \hat{\vec{E}} \left( \vec{r}{}_{\textrm{A}}, \omega, t \right) \right.\\
\left. + \hat{\vec{E}}{}^{\dagger} \left( \vec{r}_{\textrm{A}}, \omega, t \right) \sprod \left[ \vec{d}{}_{nk} \hat{A}{}_{mk} \left( t \right) - \vec{d}{}_{km} \hat{A}{}_{kn} \left( t \right) \right] \right\}.
\label{eq:Atomic Flip Operator}
\end{multline}
The electric field \eqref{eq:Electric Field Free and Induced} is evaluated using the Markov approximation for weak atom-field coupling and we discard slow non-oscillatory dynamics of the flip operator by setting $\hat{A}{}_{mn} \left( t' \right) \simeq \me^{\mi \tilde{\omega}{}_{mn} \left( t'-t \right)} \hat{A}{}_{mn} \left( t \right)$ for the time interval $t_0 \leq t' \leq t$. The dynamics is determined by the shifted frequency $\tilde{\omega}{}_{mn} = \omega_{mn} + \delta \omega_{mn}$ with the pure atom's eigenfrequency $\omega_{mn}$ and the Casimir--Polder frequency shift $\delta \omega_{mn}$ due to the presence of the surface, which is computed in the following. We make use of the relation
\begin{equation}
\int\limits^t_{-\infty}{\dif t' \me^{\pm \mi \left( \omega - \tilde{\omega}{}_{nm} \right) \left( t-t' \right)}} = \pi \delta \left( \omega - \tilde{\omega}{}_{nm} \right) \pm \mi \mathcal{P} \left( \frac{1}{\omega - \tilde{\omega}{}_{nm}} \right)
\label{eq:Delta Function}
\end{equation}
with the Cauchy principle value $\mathcal{P}$ and used $\tilde{\omega}{}_{nm} = -\tilde{\omega}{}_{mn}$. Moreover we have set the lower integral boundary from $t_0$ to infinity. The Markov approximation reduces the memory of the atomic flip operator from its entire past to present time $t$ only. To apply the Markov approximation we have assumed that the atomic transition frequency $\tilde{\omega}_{10}$ is not close to any narrow-band resonance mode of the medium. If there were such an active mode, the atom would mostly interact with it, similar to a cavity. In this case the mode would have to be modeled by a Lorentzian profile \cite{Haroche:1991, Buhmann_Welsch:2008, Buhmann_Book_2}.\\
After defining the coefficient
\begin{multline}
\vec{C}{}_{mn} = \frac{\mu_0}{\hbar} \Theta \left( \tilde{\omega}{}_{nm} \right) \tilde{\omega}{}^2{}_{nm} \mathrm{Im} \tens{G} \left( \vec{r}{}_{\textrm{A}}, \vec{r}{}_{\textrm{A}}, \tilde{\omega}{}_{nm} \right) \sprod \vec{d}{}_{mn}\\
-\frac{\mi \mu_0}{\pi \hbar} \mathcal{P} \int\limits^{\infty}_0{\dif \omega \frac{1}{\omega - \tilde{\omega}{}_{nm}} \omega^2 \mathrm{Im} \tens{G} \left( \vec{r}{}_{\textrm{A}}, \vec{r}{}_{\textrm{A}}, \omega \right) \sprod \vec{d}{}_{mn}}
\end{multline}
the equation of motion for the atomic flip operator \eqref{eq:Atomic Flip Operator} reads
\begin{multline}
\dot{\hat{A}}{}_{mn} \left( t \right) = \mi \omega_{mn} \hat{A}{}_{mn} \left( t \right)\\
+ \frac{\mi}{\hbar} \sum\limits_k \int\limits^{\infty}_0 \dif \omega \left\{ \me^{-\mi \omega \left( t-t_0 \right)} \left[ \hat{A}{}_{mk} \left( t \right) \vec{d}{}_{nk} - \hat{A}{}_{kn} \left( t \right) \vec{d}{}_{km} \right] \sprod \hat{\vec{E}} \left( \vec{r}{}_{\textrm{A}}, \omega \right) \right.\\
\left. + \me^{\mi \omega \left( t-t_0 \right)} \hat{\vec{E}}{}^{\dagger} \left( \vec{r}{}_{\textrm{A}}, \omega \right) \sprod \left[ \vec{d}{}_{nk} \hat{A}{}_{mk} \left( t \right) - \vec{d}{}_{km} \hat{A}{}_{kn} \left( t \right) \right] \right\}\\
- \sum\limits_{k,l} \left[ \vec{d}{}_{nk} \sprod \vec{C}{}_{kl} \hat{A}{}_{ml} \left( t \right) - \vec{d}{}_{km} \sprod \vec{C}{}_{nl} \hat{A}{}_{kl} \left( t \right) \right]\\
+ \sum\limits_{k,l} \left[ \vec{d}{}_{nk} \sprod \vec{C}{}^*{}_{ml} \hat{A}{}_{lk} \left( t \right) - \vec{d}{}_{km} \sprod \vec{C}{}^*{}_{kl} \hat{A}{}_{ln} \left( t \right) \right].
\end{multline}
Equations of motion for the diagonal and nondiagonal atomic flip operators can be decoupled by assuming that the atom does not have quasi-degenerate transitions. Moreover the atom is unpolarized in each of its energy eigenstates, $\vec{d}{}_{mm} = \vec{0}$, which is guaranteed by atomic selection rules \cite{Buhmann_Book_2}. Thus the fast-oscillating nondiagonal parts $\hat{A}{}_{mn} \left( t \right)$ can be decoupled from the slowly-oscillating diagonal operator terms $\hat{A}{}_{mm} \left( t \right)$. In the following we take the expectation value of the atomic flip operator \eqref{eq:Atomic Flip Operator}. By making use of the definitions of the surface-induced frequency shift and decay rate
\begin{align}
\begin{array}{lll}
&\delta \omega_{nk} &= - \frac{\mu_0}{\pi \hbar} \mathcal{P} \int\limits^{\infty}_0{\dif \omega \frac{1}{\omega - \tilde{\omega}_{nk}} \omega^2 \vec{d}_{nk} \cdot \mathrm{Im} \tens{G} ^{(1)} \left( \vec{r}_{\textrm{A}}, \vec{r}_{\textrm{A}}, \omega \right) \cdot \vec{d}_{kn}}\\
&\Gamma_{nk} &= \frac{2 \mu_0}{\hbar} \tilde{\omega}^2_{nk} \vec{d}_{nk} \cdot \mathrm{Im} \tens{G} \left( \vec{r}_{\textrm{A}}, \vec{r}_{\textrm{A}}, \tilde{\omega}_{nk} \right) \cdot \vec{d}_{kn}
\end{array}
\end{align}
the relations
\begin{align}
\begin{array}{lll}
&\delta \omega_n &= \sum\limits_k{\delta \omega_{nk}}\\[4mm]
&\Gamma_n &= \sum\limits_{k<n}{\Gamma_{nk}}
\end{array}
\end{align}
and the expression for the shifted frequency
\begin{equation}
\tilde{\omega}_{mn} = \omega_{mn} + \delta \omega_m - \delta \omega_n,
\end{equation}
the expressions for the diagonal elements and the nondiagonal elements of the atomic flip operator for a coherent electric field $\vec{E} \left( \vec{r}_{\textrm{A}}, t \right)$ \eqref{eq:Electric Driving Field} read
\begin{multline}
\langle \dot{\hat{A}}_{mm} \left( t \right) \rangle = - \Gamma_m \langle \hat{A}_{mm} \left( t \right) \rangle + \sum\limits_{k>m}{\Gamma_{km} \langle \hat{A}_{kk} \left( t \right) \rangle}\\
+\frac{\mi}{\hbar} \sum\limits_k{\left[ \langle \hat{A}_{mk} \left( t \right) \rangle \vec{d}_{mk} - \langle \hat{A}_{km} \left( t \right) \rangle \vec{d}_{km} \right] \cdot \vec{E} \left( \vec{r}_{\textrm{A}}, t \right)}
\label{eq:Atomic Flip Operator Diagonal}
\end{multline}
and
\begin{multline}
\langle \dot{\hat{A}}_{mn} \left( t \right) \rangle = \mi \tilde{\omega}_{mn} \langle \hat{A}_{mn} \left( t \right) \rangle - \frac{1}{2} \left[ \Gamma_n + \Gamma_m \right] \langle \hat{A}_{mn} \left( t \right) \rangle\\
+\frac{\mi}{\hbar} \sum\limits_k{ \left[ \langle \hat{A}_{mk} \left( t \right) \rangle \vec{d}_{nk} - \langle \hat{A}_{kn} \left( t \right) \rangle \vec{d}_{km} \right] \cdot \vec{E} \left( \vec{r}_{\textrm{A}}, t \right)}.
\label{eq:Atomic Flip Operator Nondiagonal}
\end{multline}
Whereas the diagonal terms of the atomic flip operator represent the probabilities of the atom to be in the respective state, the equation for the nondiagonal elements of the atomic flip operator are needed to compute the dipole moment \eqref{eq:Dipole Moment Definition}. As the electric field \eqref{eq:Electric Field Free and Induced} consists of two contributions, the dipole moment can also be decomposed into a free part $\hat{\vec{d}}_{\textrm{free}} \left( t \right)$ stemming from the first term in Eq.~\eqref{eq:Atomic Flip Operator Nondiagonal}, which is the homogenous solution, and the induced term $\hat{\vec{d}}_{\textrm{ind}} \left( t \right)$ from the inhomogeneous solution containing the electric field
\begin{equation}
\hat{\vec{d}} \left( t \right) = \hat{\vec{d}}_{\textrm{ind}} \left( t \right) + \hat{\vec{d}}_{\textrm{free}} \left( t \right).
\label{eq:Dipole Moment Free and Induced}
\end{equation}
This notation is schematic because the equation for the atomic flip operator containing phenomenological damping constants is only defined as an averaged quantity.\\
In the next section, Sec.~\ref{sec:Perturbative Approach}, the dipole moment is computed in a perturbative approach.

\section{Perturbative Approach for the Dipole Moment and the Electric Field}
\label{sec:Perturbative Approach}
Making use of lowest order perturbation theory, the induced part of the dipole moment $\hat{\vec{d}}_{\textrm{ind}}$ \eqref{eq:Dipole Moment Free and Induced} only depends on the free electric field $\hat{\vec{E}}_{\textrm{free}}$ \eqref{eq:Electric Field Free and Induced} and the induced electric field $\hat{\vec{E}}_{\textrm{ind}}$ is computed by using the free dipole moment $\hat{\vec{d}}_{\textrm{free}}$ only, respectively.\\
The expectation value of the dipole moment operator $\langle \hat{\vec{d}} \left( t \right) \rangle$ equals $0$, if the atom's initial state is an incoherent superposition of energy eigenstates. In this approach the atom stays in its initial state $\ket{n}$ with $\langle \hat{A}_{kl} \left( t' \right) \rangle \approx \langle \hat{A}_{kl} \left( t \right) \rangle \approx \delta_{kn} \delta_{ln}$. The expectation value of the dipole moment in the energy eigenstate $\ket{n}$ is given by the equation
\begin{equation}
\langle \hat{\vec{d}} \left( t \right) \rangle_n = \sum\limits_{k \neq n} \left[ \langle \hat{A}_{nk} \left( t \right) \rangle_n \vec{d}_{nk} + \langle \hat{A}_{kn} \left( t \right) \rangle_n \vec{d}_{kn} \right].
\label{eq:Dipole Moment General}
\end{equation}
Since the dipole moment in time domain for an atom in an energy eigenstate $\ket{n}$ only contains off-diagonal atomic flip operators, we only need the solution of the nondiagonal atomic flip operator elements \eqref{eq:Atomic Flip Operator Nondiagonal}. Moreover the free part of this term $\hat{A}_{mn}$ vanishes because of the initial condition of off-diagonal terms $ \langle \hat{A}_{mn} \left( t_0 \right) \rangle = 0$. We call the dipole moment in eigenstate $\ket{n}$ \eqref{eq:Dipole Moment General} under the influence of a coherent electric driving field \eqref{eq:Electric Driving Field} the induced dipole moment and it reads in the Markov approximation, where we set $t_0 \rightarrow - \infty$
\begin{multline}
\langle \hat{\vec{d}}_{\textrm{ind}} \left( t \right) \rangle_n = \frac{\mi}{\hbar} \sum\limits_k \int\limits^t_{-\infty} \dif t' \left\{ \me^{ \left[ \mi \tilde{\omega}_{nk} - \frac{1}{2} \left( \Gamma_n + \Gamma_k \right) \right] \left( t-t' \right)} \vec{d}_{nk} \vec{d}_{kn} \right.\\
\left.-\me^{ \left[ \mi \tilde{\omega}_{kn} - \frac{1}{2} \left( \Gamma_k + \Gamma_n \right) \right] \left( t-t' \right)} \vec{d}_{kn} \vec{d}_{nk} \right\} \cdot \vec{E} \left( \vec{r}_{\textrm{A}}, t' \right).
\label{eq:Dipole Moment Time Domain First}
\end{multline}
After inserting the electric field \eqref{eq:Electric Driving Field} into this equation and identifying the complex atomic polarizability as
\begin{multline}
\alpha_n \left( \vec{r}_{\textrm{A}}, \omega \right) = \frac{1}{\hbar} \sum\limits_k \left[ \frac{\vec{d}_{nk} \vec{d}_{kn}}{\tilde{\omega}_{kn} - \omega -\frac{\mi}{2} \left( \Gamma_n + \Gamma_k \right)} \right.\\
\left. + \frac{\vec{d}_{kn} \vec{d}_{nk}}{\tilde{\omega}_{kn} + \omega + \frac{\mi}{2} \left( \Gamma_n + \Gamma_k \right)} \right]
\label{eq:Atomic Polarizability}
\end{multline}
with the property $\alpha^*_n \left( \omega \right) = \alpha_n \left( - \omega^* \right)$ the dipole moment in time-domain reads
\begin{multline}
\langle \hat{\vec{d}}_{\textrm{ind}} \left( t \right) \rangle_n = \frac{1}{2} \left[ \alpha_n \left( \omega_{\textrm{L}} \right) \sprod \vec{E} \left( \vec{r}_{\textrm{A}} \right) \me^{-\mi \omega_{\textrm{L}}t} \right.\\
\left. + \alpha^*_n \left( \omega_{\textrm{L}} \right) \sprod \vec{E} \left( \vec{r}_{\textrm{A}} \right) \me^{\mi \omega_{\textrm{L}}t} \right].
\label{eq:Dipole Moment Time Domain}
\end{multline}
The dipole moment in frequency domain is obtained by a Fourier transform of Eq.~\eqref{eq:Dipole Moment Time Domain First}
\begin{multline}
\langle \hat{\vec{d}}_{\textrm{ind}} \left( \omega \right) \rangle_n = \frac{1}{2 \pi} \int\limits^{\infty}_{-\infty}{\dif t \me^{\mi \omega t} \langle \hat{\vec{d}}_{\textrm{ind}} \left( t \right) \rangle_n}\\
= \frac{\mi}{2 \pi \hbar} \sum\limits_k \int\limits^{\infty}_{-\infty} \dif t \me^{\mi \omega t} \int^t_{-\infty} \dif t' \left\{ \me^{\left[ \mi \tilde{\omega}_{nk} - \frac{1}{2} \left( \Gamma_m + \Gamma_n \right) \right] \left( t-t' \right)} \vec{d}_{nk} \vec{d}_{kn} \right.\\
\left. -\me^{\left[ \mi \tilde{\omega}_{kn} - \frac{1}{2} \left( \Gamma_m + \Gamma_n \right) \right] \left( t-t' \right)} \vec{d}_{kn} \vec{d}_{nk} \right\} \sprod \vec{E} \left( \vec{r}_{\textrm{A}}, t' \right).
\label{eq:Dipole Moment Fourier Transformation}
\end{multline}
By making use of the definition of the electric driving field \eqref{eq:Electric Driving Field} and \eqref{eq:Frequency Components Driving Field} and the $\delta$-function
\begin{equation}
\frac{1}{2 \pi} \int^{\infty}_{-\infty}{\dif t \me^{\mi \left( \omega - \omega_{\textrm{L}} \right) t}} = \delta \left( \omega - \omega_{\textrm{L}} \right)
\end{equation}
or simply by using the result for the dipole moment in time domain \eqref{eq:Dipole Moment Time Domain} the frequency component of the dipole moment $\langle \hat{\vec{d}}_{\textrm{ind}} \left( \omega \right) \rangle_n$ \eqref{eq:Dipole Moment Fourier Transformation} can be written as
\begin{equation}
\langle \hat{\vec{d}}_{\textrm{ind}} \left( \omega \right) \rangle_n = \alpha_n \left( \omega_{\textrm{L}} \right) \sprod \vec{E} \left( \vec{r}_{\textrm{A}}, \omega  \right).
\label{eq:Dipole Moment Frequency Domain}
\end{equation}
We have discarded the $\delta \left( \omega + \omega_{\textrm{L}} \right)$-terms in \eqref{eq:Dipole Moment Frequency Domain} which do not contribute in the reverse transformation of the dipole component from frequency domain to time-domain, which is given by
\begin{equation}
\langle \hat{\vec{d}}_{\textrm{ind}} \left( t \right) \rangle_n = \int\limits^{\infty}_0{\dif \omega \left[ \me^{-\mi \omega t} \langle \hat{\vec{d}}_{\textrm{ind}} \left( \omega \right) \rangle_n + \me^{\mi \omega t} \langle \hat{\vec{d}}^{\dagger}_{\textrm{ind}} \left( \omega \right) \rangle_n \right]}.
\label{eq:Back Transformation}
\end{equation}
In the next step, we insert the induced dipole moment \eqref{eq:Dipole Moment Time Domain} back into the expression of the induced electric field \eqref{eq:Electric Field Free and Induced} to calculate a higher order term of the induced electric field yielding
\begin{multline}
\langle \hat{\vec{E}}{}^{(2)}_{\textrm{ind}} \left( \vec{r}, t \right) \rangle = \frac{\mi \mu_0}{\pi} \int\limits^{\infty}_0 \dif \omega \omega^2\\
\times \left\{ \frac{1}{2} \me^{-\mi \omega_{\textrm{L}} t} \left[ \pi \delta \left( \omega - \omega_{\textrm{L}} \right) - \mi \mathcal{P} \frac{1}{\omega - \omega_{\textrm{L}}} \right] \right.\\
\times \mathrm{Im} \tens{G} \left( \vec{r}, \vec{r}_{\textrm{A}}, \omega \right) \sprod \alpha_n \left( \omega_{\textrm{L}} \right) \sprod \vec{E} \left( \vec{r}_{\textrm{A}} \right)\\[2mm]
+ \frac{1}{2} \me^{\mi \omega_{\textrm{L}} t} \left[ \pi \delta \left( \omega + \omega_{\textrm{L}} \right) - \mi \mathcal{P} \frac{1}{\omega + \omega_{\textrm{L}}} \right]\\
\times \mathrm{Im} \tens{G} \left( \vec{r}, \vec{r}_{\textrm{A}}, \omega \right) \sprod \alpha^*_n \left( \omega_{\textrm{L}} \right) \sprod \vec{E} \left( \vec{r}_{\textrm{A}} \right)\\[2mm]
- \frac{1}{2} \me^{\mi \omega_{\textrm{L}} t} \left[ \pi \delta \left( \omega - \omega_{\textrm{L}} \right) + \mi \mathcal{P} \frac{1}{\omega - \omega_{\textrm{L}}} \right]\\
\times \mathrm{Im} \tens{G} \left( \vec{r}, \vec{r}_{\textrm{A}}, \omega \right) \sprod \alpha^*_n \left( \omega_{\textrm{L}} \right) \sprod \vec{E} \left( \vec{r}_{\textrm{A}} \right)\\[2mm]
- \frac{1}{2} \me^{-\mi \omega_{\textrm{L}} t} \left[ \pi \delta \left( \omega + \omega_{\textrm{L}} \right) + \mi \mathcal{P} \frac{1}{\omega + \omega_{\textrm{L}}} \right]\\
\left. \times \mathrm{Im} \tens{G} \left( \vec{r}, \vec{r}_{\textrm{A}}, \omega \right) \sprod \alpha_n \left( \omega_{\textrm{L}} \right) \sprod \vec{E} \left( \vec{r}_{\textrm{A}} \right) \right\}.
\end{multline}
The expressions containing $\delta \left( \omega + \omega_{\textrm{L}} \right)$ do not contribute to the electric field under the integration over $\omega$ from $0$ to $\infty$. One can make use of the definition of the imaginary part of the Green's tensor
\begin{equation}
\mathrm{Im} \tens{G} \left( \vec{r}, \vec{r}_{\textrm{A}}, \omega \right) = \frac{1}{2 \mi} \left[ \tens{G} \left( \vec{r}, \vec{r}_{\textrm{A}}, \omega \right) - \tens{G}^* \left( \vec{r}, \vec{r}_{\textrm{A}}, \omega \right) \right]
\end{equation}
and the Schwarz' principle $\tens{G}^* \left( \vec{r}, \vec{r}_{\textrm{A}}, \omega \right) = \tens{G} \left( \vec{r}, \vec{r}_{\textrm{A}}, -\omega^* \right)$ for real frequencies $\omega = \omega^*$. The integrals over the Cauchy principle value $\mathcal{P} \int^{\infty}_0{\dif \omega/ \left( \omega - \omega_{\textrm{L}} \right)}$ and $\mathcal{P} \int^{-\infty}_0{\dif \omega/ \left( \omega + \omega_{\textrm{L}} \right)}$ have poles along the curve of integration at $\omega_{\textrm{L}}$ and $-\omega_{\textrm{L}}$, respectively, and are evaluated in the complex plane. There is a part along the quarter circle, which vanishes because of $\lim_{|\omega| \rightarrow 0} \tens{G}^{(1)} \left( \vec{r}, \vec{r}_{\textrm{A}}, \omega \right) \omega^2/c^2 = 0$. The Green's tensor is evaluated at complex frequencies $\omega \rightarrow \mi \xi$ in the part along the imaginary frequency axis. Expressions containing discrete frequencies $\omega_{\textrm{L}}$ are obtained by computing the poles. The integrals $\mathcal{P} \int^{\infty}_0{\dif \omega / \left( \omega + \omega_{\textrm{L}} \right)}$ and $\mathcal{P} \int^{-\infty}_0{\dif \omega / \left( \omega - \omega_{\textrm{L}} \right)}$ do not contain poles. Their contributions are thus equal to the part along the imaginary axis.\\
After bringing together the calculations from all parts, the nonresonant part stemming from the integration along the imaginary frequency axis vanishes at all and only a resonant contribution containing discrete frequencies $\omega_{\textrm{L}}$
\begin{multline}
\langle \hat{\vec{E}}^{(2)}_{\textrm{ind}} \left( \vec{r}, t \right) \rangle = \frac{1}{2} \mu_0 \omega^2_{\textrm{L}} \me^{-\mi \omega_{\textrm{L}} t} \tens{G} \left( \vec{r}, \vec{r}_{\textrm{A}}, \omega_{\textrm{L}} \right) \sprod \alpha_n \left( \omega_{\textrm{L}} \right) \sprod \vec{E} \left( \vec{r}_{\textrm{A}} \right)\\
+ \frac{1}{2} \mu_0 \omega^2_{\textrm{L}} \me^{\mi \omega_{\textrm{L}} t} \tens{G}^* \left( \vec{r}, \vec{r}_{\textrm{A}}, \omega_{\textrm{L}} \right) \sprod \alpha_n \left( \omega_{\textrm{L}} \right) \sprod \vec{E} \left( \vec{r}_{\textrm{A}} \right).
\label{eq:Electric Field Induced}
\end{multline}
remains. The final expression for the electric field shows the Green's tensor and the atomic polarizability at the laser frequency $\omega_{\textrm{L}}$. The final expression for the electric field shows the Green's tensor and the atomic polarizability at the laser frequency $\omega_{\textrm{L}}$. The shifted atomic transition frequency $\tilde{\omega}_{nk}$ only enters the expression as part of the atomic polarizability in Eq.~\eqref{eq:Atomic Polarizability}. This time-dependent expression shows oscillations with the laser frequency $\omega_{\textrm{L}}$ as opposed to the atomic transition frequency in the term of lowest order \eqref{eq:Electric Field Induced}. Moreover, there is a scaling of the electric field emitted by the atom with the amplitude of the electric driving field $\vec{E} \left( \vec{r}_{\textrm{A}} \right)$. This opens up the possibility to enhance the electric field emitted by an atom by increasing the laser intensity.\\
The higher order result for the induced electric field \eqref{eq:Electric Field Induced} can be inserted into the equation of the induced dipole moment \eqref{eq:Dipole Moment Time Domain First} leading to a higher order induced dipole moment. By dropping the counter-rotating terms, the higher order dipole moment reads
\begin{multline}
\langle \hat{\vec{d}}_{\textrm{ind}}^{(4)} \left( t \right) \rangle =\\
\frac{1}{2} \mu_0 \omega^2_{\textrm{L}}  \me^{-\mi \omega_{\textrm{L}} t} \alpha_n \left( \omega_{\textrm{L}} \right) \sprod \tens{G} \left( \vec{r}_{\textrm{A}}, \vec{r}_{\textrm{A}}, \omega_{\textrm{L}} \right) \sprod \alpha_n \left( \omega_{\textrm{L}} \right) \sprod \vec{E} \left( \vec{r}_{\textrm{A}} \right)\\
+\frac{1}{2} \mu_0 \omega^2_{\textrm{L}} \me^{\mi \omega_{\textrm{L}} t} \alpha^*_n \left( \omega_{\textrm{L}} \right) \sprod \tens{G}^* \left( \vec{r}_{\textrm{A}}, \vec{r}_{\textrm{A}}, \omega_{\textrm{L}} \right) \sprod \alpha^*_n \left( \omega_{\textrm{L}} \right) \sprod \vec{E} \left( \vec{r}_{\textrm{A}} \right).
\label{eq:Dipole Moment Higher Order}
\end{multline}
The order is determined by the number of dipole moments in the respective expression. The higher order results of both the induced electric field \eqref{eq:Electric Field Induced} and the induced dipole moment \eqref{eq:Dipole Moment Higher Order} are inserted into an expression of the elctromagnetic potential in the following Sec.~\ref{sec:Components of the Electric Potential}.

\section{Components of the Electromagnetic Potential}
\label{sec:Components of the Electric Potential}
\begin{figure}[!ht]
\centerline{\includegraphics[width=0.7\columnwidth]{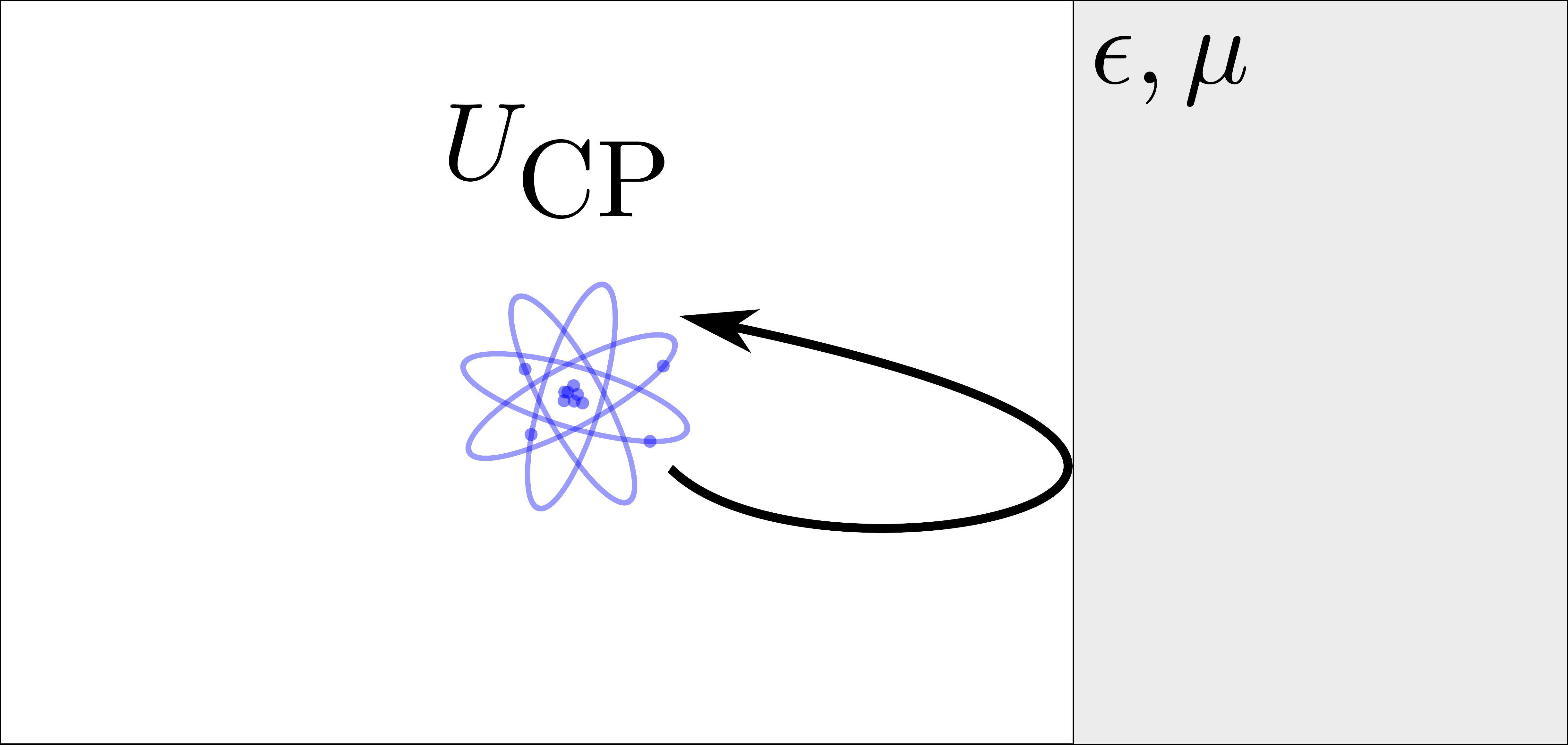}}
\caption{Sketch of the atom close to the surface. The dipole fluctuations cause the undriven Casimir--Polder potential $U_{\textrm{CP}}$.}
\label{fig:Atom_Surface}
\end{figure}

\begin{figure}[!ht]
\centerline{\includegraphics[width=0.7\columnwidth]{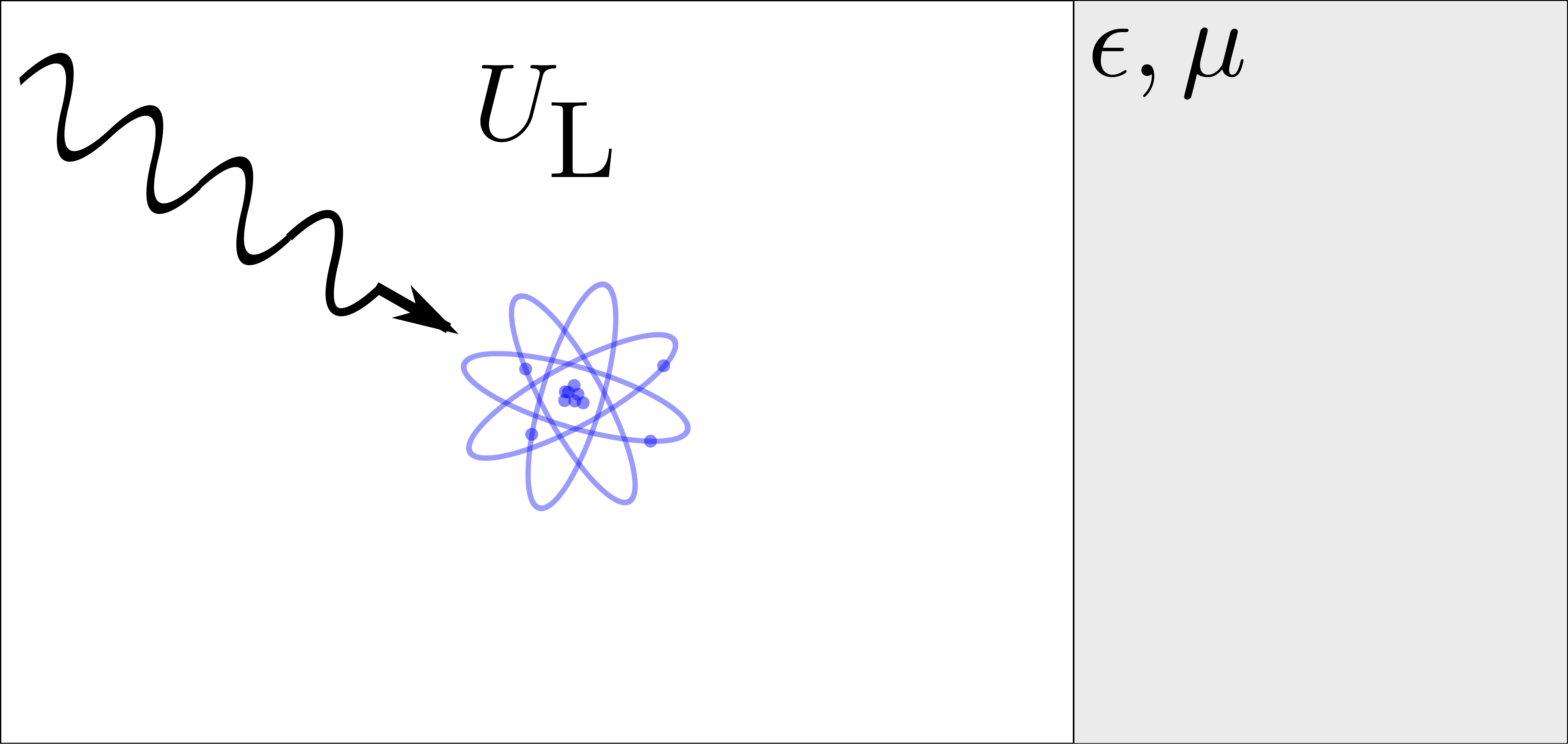}}
\caption{Sketch of the atom under the influence of the laser field. The laser field leads to the occurence of the laser light potential $U_{\textrm{L}}$.}
\label{fig:Atom_Laser}
\end{figure}

\begin{figure}[!ht]
\centerline{\includegraphics[width=0.7\columnwidth]{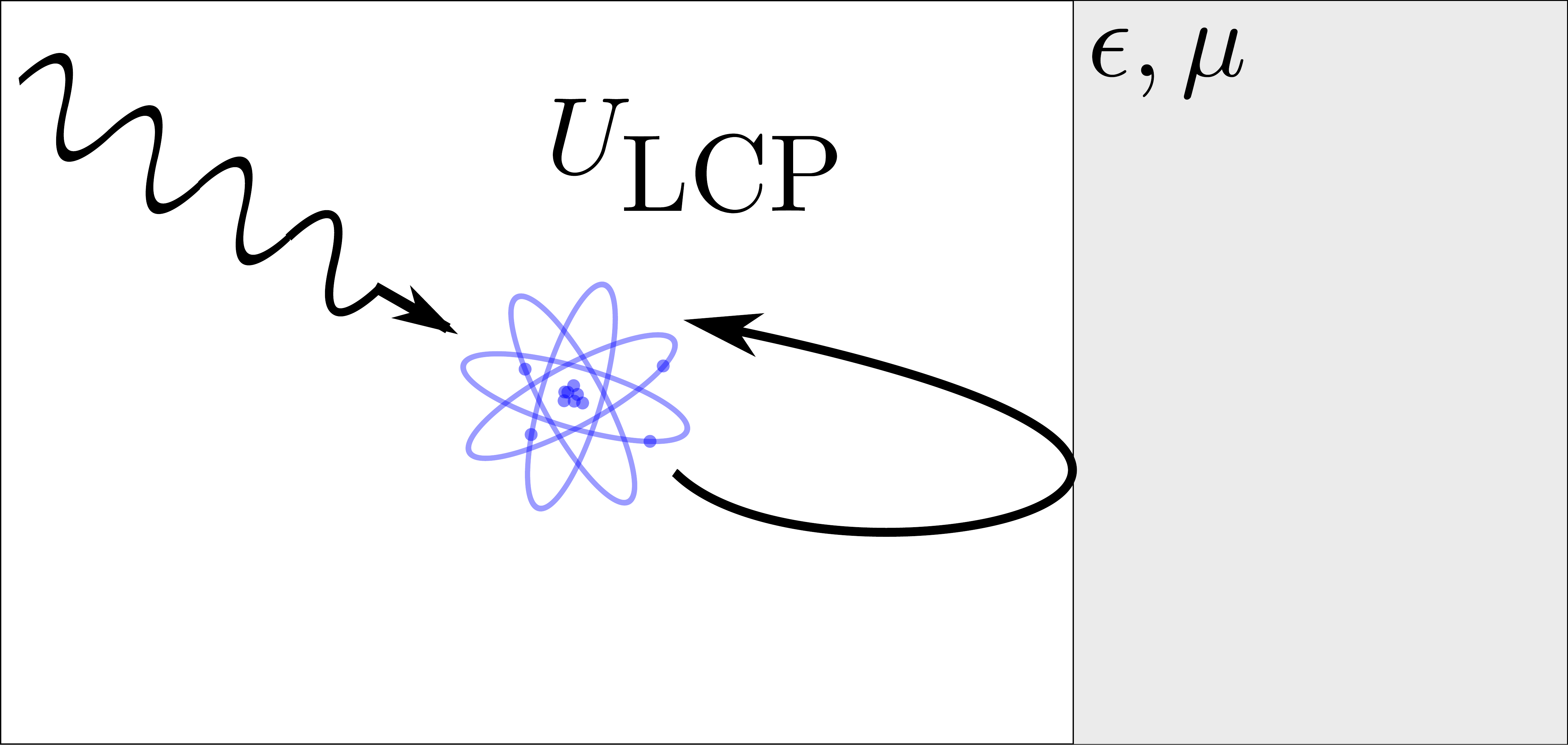}}
\caption{Sketch of the atom under the influence of the laser field and the vacuum fluctuations leading to the potential term $U_{\textrm{LCP}}$.}
\label{fig:Atom_Laser_Surface}
\end{figure}

Both the electric field \eqref{eq:Electric Field Free and Induced} and the dipole moment \eqref{eq:Dipole Moment Definition} can be decomposed into a spontaneously fluctuating free part and an induced contribution. 
Since the distance between the atom and the laser is assumed to be large, there is no back-action from the atom to the laser. The general expression for the potential is a combination of all of these contributions and reads in normal ordering (as indicated by : ... :)
\begin{align}
\begin{array}{lll}
&U \left( \vec{r}_{\textrm{A}}, t \right) &= - \displaystyle{\frac{1}{2}} \langle \hat{\vec{d}} \left( t \right) \sprod \hat{\vec{E}} \left(  \vec{r}_{\textrm{A}}, t \right) \rangle\\[3mm]
=& - \displaystyle{\frac{1}{2}} \langle : \hat{\vec{d}}_{\textrm{free}} \left( t \right) \sprod \hat{\vec{E}}_{\textrm{free}} \left(  \vec{r}_{\textrm{A}}, t \right) : \rangle &- \displaystyle{\frac{1}{2}} \langle : \hat{\vec{d}}_{\textrm{free}} \left( t \right) \sprod \hat{\vec{E}}_{\textrm{ind}} \left(  \vec{r}_{\textrm{A}}, t \right) : \rangle\\[3mm]
&- \displaystyle{\frac{1}{2}} \langle : \hat{\vec{d}}_{\textrm{ind}} \left( t \right) \sprod \hat{\vec{E}}_{\textrm{free}} \left(  \vec{r}_{\textrm{A}}, t \right) : \rangle &- \displaystyle{\frac{1}{2}} \langle : \hat{\vec{d}}_{\textrm{ind}} \left( t \right) \sprod \hat{\vec{E}}_{\textrm{ind}} \left(  \vec{r}_{\textrm{A}}, t \right) : \rangle
\label{eq:Potential Terms}
\end{array}
\end{align}
giving rise to four different terms that are analyzed in the following. This decomposition of the Casimir--Polder potential was carried out in Refs.~\cite{Dalvit_Book, Henkel:2002}. We incorporate an additional contribution of the electric field from the coherent driving laser field \eqref{eq:Electric Field Quantum State}. The first term contains the free dipole moment and the free electric field and therefore is of lowest order in perturbation theory. For $\vec{r} \; ' \in V_{\textrm{S}}$, this expression leads to the vanishing expectation value of the free dipole moment $\langle \hat{\vec{d}}_{\textrm{free}} \left( t \right) \rangle = 0$. In case of $\vec{r} \; ' \notin V_{\textrm{S}}$, this term vanishes as well according to Eqs.~\eqref{eq:Quantum State} and \eqref{eq:Electric Field Quantum State}.\\
The second term inserts the free dipole moment into the induced electric field \eqref{eq:Electric Field Free and Induced}. The expression
\begin{multline}
U_{\textrm{CP}} \left( \vec{r}_{\textrm{A}}, t \right) = -\frac{1}{2} \langle : \hat{\vec{d}}_{\textrm{free}} \left( t \right) \sprod \hat{\vec{E}}_{\textrm{ind}} \left( \vec{r}_{\textrm{A}}, t \right) : \rangle\\
= - \frac{\mi \mu_0}{2 \pi} \int\limits^{\infty}_0 \omega^2 \int\limits^t_{0} \dif t' \me^{-\mi \omega \left( t-t' \right)}\\
\langle \hat{\vec{d}}_{\textrm{free}} \left( t \right) \sprod \mathrm{Im} \tens{G} \left( \vec{r}_{\textrm{A}}, \vec{r}{}_{\textrm{A}}, \omega \right) \sprod \hat{\vec{d}}_{\textrm{free}} \left( t' \right) \rangle\\
+ \textrm{h.c.}
\end{multline}
is the standard Casimir--Polder potential, is called radiation-reaction term and stems from the dipole fluctuations \cite{Dalvit_Book}. Fig.~\ref{fig:Atom_Surface} shows a sketch of the standard Casimir--Polder potential.
The third term in Eq.~\eqref{eq:Potential Terms} has a contribution from the coherent electric field $\left( \vec{r} \; '\in V_{\textrm{S}} \right)$ and is identified with the light force of the laser on the atom. Fig.~\ref{fig:Atom_Laser} shows a sketch of the interaction between the atom and the laser field.
The electric driving field causes a force on the atom, which is associated with the AC Stark shift. This laser light potential is defined as
\begin{align}
\begin{array}{lll}
&U_{\textrm{L}} \left( \vec{r}_{\textrm{A}}, t \right) &= - \displaystyle{\frac{1}{2}} \langle : \hat{\vec{d}}_{\textrm{ind}} \left( t \right) \sprod \hat{\vec{E}}_{\textrm{free}} \left( \vec{r}_{\textrm{A}}, t \right) : \rangle\\[2mm]
& &= - \displaystyle{\frac{1}{2}} \langle \hat{\vec{d}}_{\textrm{ind}} \left( t \right) \rangle_n \sprod \vec{E} \left( \vec{r}_{\textrm{A}},t \right).
\label{eq:Force Free Electric Laser Field}
\end{array}
\end{align}
By inserting the result for the expectation value of the dipole moment operator in eigenstate $\ket{n}$ in time domain \eqref{eq:Dipole Moment Time Domain} and the electric driving field \eqref{eq:Electric Driving Field} we obtain the result
\begin{equation}
U_{\textrm{L}} \left( \vec{r}_{\textrm{A}}, t \right) = - \frac{1}{2} \vec{E} \left( \vec{r}_{\textrm{A}} \right) \sprod \alpha_n \left( \omega_{\textrm{L}} \right) \sprod \vec{E} \left( \vec{r}_{\textrm{A}} \right) \cos^2 \left( \omega_{\textrm{L}} t \right).
\end{equation}
Under the assumption of real polarizabilities with real dipole moments $\vec{d}_{10} = \vec{d}_{01} = \vec{d}$ for a two-level atom with a transition frequency $\tilde{\omega}_{10}$ and no damping, the atomic polarizability reads
\begin{equation}
\alpha \left( \omega \right) = \frac{1}{\hbar} \left[ \frac{\vec{d} \vec{d}}{ \tilde{\omega}_{10} - \omega_{\textrm{L}}} + \frac{\vec{d} \vec{d}}{\tilde{\omega}_{10} + \omega_{\textrm{L}}} \right] = \frac{1}{\hbar} \frac{2 \tilde{\omega}_{10} \vec{d} \vec{d}}{\tilde{\omega}^2_{10} - \omega^2_{\textrm{L}}}.
\label{eq:Atomic Polarizability10}
\end{equation}
Since the damping rates are set equal to $0$, the atomic polarizability \eqref{eq:Atomic Polarizability} is real-valued. For an isotropic atomic state it reads
\begin{equation}
\alpha \left( \omega_{\textrm{L}} \right) = \frac{2 \tilde{\omega}{}_{10} d^2}{3 \hbar \left( \tilde{\omega}{}^2_{10} - \omega^2_{\textrm{L}} \right)} \tens{1},
\label{eq:Atomic Polarizability Unitary Matrix}
\end{equation}
where the factor of $1/3$ stems from isotropy. By assuming a small detuning in comparison to the atomic transition frequency $\Delta = \omega_{\textrm{L}} - \tilde{\omega}_{10} \ll \tilde{\omega}_{10}$, which is usually guaranteed, the atomic polarizability reads
\begin{equation}
\alpha \left( \omega_{\textrm{L}} \right) \approx - \frac{d^2}{3 \hbar \Delta} \tens{1}.
\label{eq:Atomic Polarizability Detuning}
\end{equation}
The potential of the light force under these approximations is given by
\begin{equation}
U_{\textrm{L}} \left( \vec{r}_{\textrm{A}}, t \right) \approx \frac{1}{12} \frac{d^2 \vec{E}^2 \left( \vec{r}_{\textrm{A}} \right)}{\hbar \Delta},
\label{eq:Free Electric Force Perturbation}
\end{equation}
where we have averaged over fast oscillating terms.\\
The light force only depends on the field strength of the laser, the detuning between the atomic transition frequency and the laser frequency and the atomic dipole moment. Due to the direct interaction of atom and laser without taking the surface into account, there is no dependence on the distance.\\
The light force acts upon the atom depending on the atom's energy state. Weak-field seekers show an electric moment that is antialigned with the electric field so that they are attracted towards a local minimum of the magnitude of the electric laser field \cite{Cornell:1991}. In contrast to that, high-field seekers are drawn towards a local maximum in the energy landscape of the electric field. Refs.~\cite{Cornell:1991, Spreeuw:1994} describe magnetic trapping techniques for neutral atoms, where the significance of the atomic energy state for the trapping procedure is outlined.\\
The fourth term in Eq.~\eqref{eq:Potential Terms} $\langle \hat{\vec{d}}_{\textrm{ind}} \left( t \right) \sprod \hat{\vec{E}}_{\textrm{ind}} \left( \vec{r}_{\textrm{A}}, t \right) \rangle$, includes the induced dipole moment and the induced electric field and is one order higher in perturbation theory. Both for $\vec{r} \; ' \notin V_{\textrm{S}}$ and for $\vec{r} \; ' \in V_{\textrm{S}}$, using Eqs.~\eqref{eq:Quantum State} and \eqref{eq:Electric Field Quantum State}, this term reduces to $\langle \hat{\vec{d}}_{\textrm{free}} \left( t \right) \rangle = 0$ and thus vanishes.\\
After analyzing Eq.~\eqref{eq:Potential Terms}, expressions of next higher order can be set up by making use of the induced electric field \eqref{eq:Electric Field Induced} and the induced dipole moment \eqref{eq:Dipole Moment Higher Order} of higher order. The fourth term of Eq.~\eqref{eq:Potential Terms} leads to the result
\begin{multline}
- \frac{1}{2} \langle : \hat{\vec{d}}_{\textrm{ind}} \left( t \right) \sprod \hat{\vec{E}}^{(2)}_{\textrm{ind}} \left( \vec{r}_{\textrm{A}}, t \right) : \rangle\\[2mm]
= -\frac{1}{4} \mu_0 \omega^2_{\textrm{L}} \vec{E} \left( \vec{r}_{\textrm{A}} \right) \sprod \alpha_n \left( \omega_{\textrm{L}} \right) \sprod \mathrm{Re} \tens{G} \left( \vec{r}_{\textrm{A}}, \vec{r}_{\textrm{A}}, \omega_{\textrm{L}} \right) \sprod \alpha_n \left( \omega_{\textrm{L}} \right) \sprod \vec{E} \left( \vec{r}_{\textrm{A}} \right),
\label{eq:Contribution Fourth Term}
\end{multline}
where we have assumed real and isotropic polarizabilities \eqref{eq:Atomic Polarizability Unitary Matrix}.
The electric field emitted by the atom \eqref{eq:Electric Field Induced} is evaluated at the atom's position $\vec{r}{}_{\textrm{A}}$. We have discarded fast oscillating terms with $\me^{-2 \mi \omega_{\textrm{L}} t}$ and $\me^{2 \mi \omega_{\textrm{L}} t}$ so that the final result for the potential does not show a time-dependence anymore.\\
The third term in Eq.~\eqref{eq:Potential Terms}, $\langle \hat{\vec{d}}^{(2)}_{\textrm{ind}} \left( t \right) \sprod \hat{\vec{E}}_{\textrm{free}} \left( \vec{r}_{\textrm{A}}, t \right) \rangle$, is computed by using Eq.~\eqref{eq:Dipole Moment Higher Order} and gives the exact same expression as Eq.~\eqref{eq:Contribution Fourth Term} so that the total laser-driven Casimir--Polder potential eventually reads
\begin{multline}
U^{\textrm{per}}_{\textrm{LCP}} \left( \vec{r}_{\textrm{A}}, t \right)\\
= -\frac{1}{2} \langle : \hat{\vec{d}}_{\textrm{ind}} \left( t \right) \sprod \hat{\vec{E}}^{(2)}_{\textrm{ind}} \left( \vec{r}_{\textrm{A}}, t \right) : \rangle - \frac{1}{2} \langle : \hat{\vec{d}}^{(2)}_{\textrm{ind}} \left( t \right) \sprod \hat{\vec{E}}_{\textrm{free}} \left( \vec{r}_{\textrm{A}}, t \right) : \rangle\\
= -\frac{1}{2} \mu_0 \omega^2_{\textrm{L}} \vec{E} \left( \vec{r}_{\textrm{A}} \right) \sprod \alpha_n \left( \omega_{\textrm{L}} \right) \sprod \mathrm{Re} \tens{G} \left( \vec{r}_{\textrm{A}}, \vec{r}_{\textrm{A}}, \omega_{\textrm{L}} \right) \sprod \alpha_n \left( \omega_{\textrm{L}} \right) \sprod \vec{E} \left( \vec{r}_{\textrm{A}} \right).
\label{eq:Driven Casimir-Polder Potential Perturbation}
\end{multline}
Fig.~\ref{fig:Atom_Laser_Surface} shows the atom under the influence of vacuum fluctuations and the laser field.
The Casimir--Polder force corresponding to the respective potential \eqref{eq:Driven Casimir-Polder Potential Perturbation} is computed by taking the gradient of the potential
\begin{equation}
\vec{F}_{\textrm{LCP}} \left( \vec{r}_{\textrm{A}} \right) = - \overrightarrow{\nabla}_{\textrm{A}} U_{\textrm{LCP}} \left( \vec{r}_{\textrm{A}} \right)
\end{equation}
and can be expressed using the two contributions
\begin{equation}
\overrightarrow{\nabla} \langle : \hat{\vec{d}}^{(2)}_{\textrm{ind}} \sprod \hat{\vec{E}}_{\textrm{free}} \left( \vec{r} \right) : \rangle_{\vec{r} = \vec{r}_{\textrm{A}}} + \overrightarrow{\nabla} \langle : \hat{\vec{d}}_{\textrm{ind}} \sprod \hat{\vec{E}}^{(2)}_{\textrm{ind}} \left( \vec{r} \right) : \rangle_{\vec{r} = \vec{r}_{\textrm{A}}},
\end{equation}
where one can use the relation $\left. \overrightarrow{\nabla} \vec{E} \left( \vec{r}_{\textrm{A}} \right) \sprod \vec{E} \left( \vec{r} \right) \right|_{\vec{r} = \vec{r}_{\textrm{A}}} = \frac{1}{2} \overrightarrow{\nabla}_{\textrm{A}} \vec{E}^2 \left( \vec{r}_{\textrm{A}} \right)$ and the symmetry of the Green's tensor $\overrightarrow{\nabla} \tens{G} \left( \vec{r}, \vec{r}_{\textrm{A}} \right) = \frac{1}{2} \overrightarrow{\nabla}_{\textrm{A}} \tens{G} \left( \vec{r}_{\textrm{A}}, \vec{r}_{\textrm{A}} \right)$. The result \eqref{eq:Driven Casimir-Polder Potential Perturbation} is analyzed further for a special geometry and an atom in Sec.~\ref{sec:Atom Near a Plane Surface} and is compared with the findings in Ref.~\cite{Perreault:2008}.

\section{Potential Ansatz with Rabi Oscillations}
\label{sec:Bloch Equation}
In this approach, the atomic dipole moment \eqref{eq:Dipole Moment General} is also computed, but there is a strong coupling between the atom and the laser field. As a result the diagonal terms of the atomic flip operator \eqref{eq:Atomic Flip Operator Diagonal} play an important role in the internal dynamics. Both the results for the light force on the atom and the Casimir--Polder potential have to be adjusted to this case.

\subsubsection{Dynamics and Dipole Moments}
\label{sec:Dynamics and Dipole Moments}
Whereas Secs.~\ref{sec:Perturbative Approach} and \ref{sec:Components of the Electric Potential} study the internal atomic dynamics for the case where the atom stays in its initial state, this assumption is not made in this section. A strong coupling between the atom and the laser frequency manifesting itself in Rabi oscillations is assumed.\\
We look at a two-level atom and want to study the internal dynamics. In order to compute the dipole moment in time-domain \eqref{eq:Dipole Moment Time Domain} we use the non-diagonal atomic flip operator \eqref{eq:Atomic Flip Operator Nondiagonal} and obtain equations of motion for $\langle \hat{A}_{10} \left( t \right) \rangle$ by setting $m=1$ and $n=0$ and for $\langle \hat{A}_{01} \left( t \right) \rangle$, where we have set $m=0$ and $n=1$
\begin{align}
\begin{array}{llll}
&\langle \dot{\hat{A}}_{10} \left( t \right) \rangle &=& \mi \tilde{\omega}_{10} \langle \hat{A}_{10} \left( t \right) \rangle\\[2mm]
& & &+ \displaystyle{\frac{\mi}{\hbar}} \left[  \langle \hat{A}_{11} \left( t \right) \rangle - \langle \hat{A}_{00} \left( t \right) \rangle \right] \vec{d}_{01} \sprod \vec{E} \left( \vec{r}_{\textrm{A}}, t \right)\\[2mm]
&\langle \dot{\hat{A}}_{01} \left( t \right) \rangle &=& \mi \tilde{\omega}_{01} \langle \hat{A}_{01} \left( t \right) \rangle\\[2mm]
& & &+ \displaystyle{\frac{\mi}{\hbar}} \left[  \langle \hat{A}_{00} \left( t \right) \rangle - \langle \hat{A}_{11} \left( t \right) \rangle \right] \vec{d}_{10} \sprod \vec{E} \left( \vec{r}_{\textrm{A}}, t \right).
\end{array}
\end{align}
The dipole moments $\vec{d}_{00}$ and $\vec{d}_{11}$ are equal to $0$ and we have set the damping terms $\Gamma_0$ and $\Gamma_1$ to $0$. Since the nondiagonal terms $\langle \hat{A}_{10} \left( t \right) \rangle$ and $\langle \hat{A}_{01} \left( t \right) \rangle$ couple to $\langle \hat{A}_{00} \left( t \right) \rangle$ and $\langle \hat{A}_{11} \left( t \right) \rangle$, one has to use Eq.~\eqref{eq:Atomic Flip Operator Diagonal} to compute the diagonal flip operators by setting $m=1$ and $m=0$, respectively,
\begin{align}
\begin{array}{lll}
\langle \dot{\hat{A}}_{11} \left( t \right) \rangle &=&  \displaystyle{\frac{\mi}{\hbar}} \left[ \langle \hat{A}_{10} \left( t \right) \rangle \vec{d}_{10} - \langle \hat{A}_{01} \left( t \right) \rangle \vec{d}_{01} \right] \cdot \vec{E} \left( \vec{r}_{\textrm{A}}, t \right)\\[2mm]
\langle \dot{\hat{A}}_{00} \left( t \right) \rangle &=& \displaystyle{\frac{\mi}{\hbar}} \left[ \langle \hat{A}_{01} \left( t \right) \rangle \vec{d}_{01} - \langle \hat{A}_{10} \left( t \right) \rangle \vec{d}_{10} \right] \cdot \vec{E} \left( \vec{r}_{\textrm{A}}, t \right).
\end{array}
\end{align}
In the following we assume real dipole moments $\vec{d}_{10} = \vec{d}_{01} = \vec{d}$. After inserting the electric driving field \eqref{eq:Electric Driving Field}, fast oscillating terms are discarded according to the Rotating Wave Approximation (RWA). Moreover we define a frame rotating with the laser frequency $\langle \hat{\tilde{A}}_{10} \left( t\right) \rangle = \me^{-\mi \omega_{\textrm{L}} t} \langle \hat{A}_{10} \left( t \right) \rangle$ and $\langle \hat{\tilde{A}}_{01} \left( t \right) \rangle = \me^{\mi \omega_{\textrm{L}} t} \langle \hat{A}_{01} \left( t\right) \rangle$. Additionally, the Rabi frequency $\Omega$ and the detuning $\Delta$ are defined as
\begin{equation}
\Omega = \frac{\vec{E} \left( \vec{r}_{\textrm{A}} \right) \sprod \vec{d}}{\hbar}; \; \Delta = \omega_{\textrm{L}} - \tilde{\omega}_{10}.
\label{eq:Definition Frequency}
\end{equation}
The new set of equations reads
\begin{align}
\begin{array}{lllll}
&\langle \dot{\hat{\tilde{A}}}_{10} \left( t \right) \rangle &= -\mi \Delta \langle \hat{\tilde{A}}_{10}  \left( t \right) \rangle &+ \frac{1}{2} \mi \Omega  \left[ \langle \hat{A}_{11} \left( t \right) \rangle - \langle \hat{A}_{00} \left( t \right) \rangle \right]\\[2mm]
&\langle \dot{\hat{\tilde{A}}}_{01} \left( t \right) \rangle &= \mi \Delta \langle \hat{\tilde{A}}_{01} \left( t \right) \rangle &- \frac{1}{2} \mi \Omega \left[ \langle \hat{A}_{11} \left( t \right) \rangle - \langle \hat{A}_{00} \left( t \right) \rangle \right]\\[2mm]
&\langle \dot{\hat{A}}_{11} \left( t \right) \rangle &= & \frac{1}{2} \mi \Omega \left[ \langle \hat{\tilde{A}}_{10}  \left( t \right) \rangle - \langle \hat{\tilde{A}}_{01} \left( t \right) \rangle \right]\\[2mm]
&\langle \dot{\hat{A}}_{00} \left( t \right) \rangle &= & - \frac{1}{2} \mi \Omega \left[ \langle \hat{\tilde{A}}_{10} \left( t \right) \rangle - \langle \hat{\tilde{A}}_{01} \left( t \right) \rangle  \right].
\end{array}
\end{align}
This system of differential equations is solved by introducing new variables
\begin{align}
\begin{array}{lll}
&\hat{A}_+ &= \frac{1}{2} \left( \hat{A}_{11} + \hat{A}_{00} \right)\\[2mm]
&\hat{A}_- &= \frac{1}{2} \left( \hat{A}_{11} - \hat{A}_{00} \right)\\[2mm]
&\hat{A}_{\textrm{I}} &= \frac{1}{2} \left( \hat{\tilde{A}}_{10} + \hat{\tilde{A}}_{01} \right)\\[2mm]
&\hat{A}_{\textrm{II}} &= \frac{1}{2} \left( \hat{\tilde{A}}_{10} - \hat{\tilde{A}}_{01} \right)
\end{array}
\end{align}
and we consider the initial conditions $\langle \hat{A}_{00} \left( 0 \right) \rangle = 1$ and $\langle \hat{A}_{11} \left( 0 \right) \rangle = \langle \hat{A}_{01} \left( 0 \right) \rangle = \langle \hat{A}_{10} \left( 0 \right) \rangle = 0$. The final solution of the system of equations of motion reads
\begin{align}
\begin{array}{llll}
&\langle \hat{A}_{00} \left( t \right) \rangle &=& \displaystyle{\frac{\Omega^2}{\Delta^2 + \Omega^2}} \cos^2 \left( \frac{1}{2} \sqrt{ \Delta^2 + \Omega^2 } t \right) + \frac{\Delta^2}{\Delta^2 + \Omega^2}\\[3mm]
&\langle \hat{A}_{11} \left( t \right) \rangle &=& \displaystyle{\frac{\Omega^2}{\Delta^2 + \Omega^2}} \sin^2 \left( \frac{1}{2} \sqrt{\Delta^2 + \Omega^2} t \right)\\[3mm]
&\langle \hat{A}_{10} \left( t \right) \rangle &=& -\displaystyle{\frac{\Omega \Delta}{\Delta^2 + \Omega^2}}\sin^2 \left( \frac{1}{2} \sqrt{\Delta^2 + \Omega^2} t \right) \me^{\mi \omega_{\textrm{L}} t}\\[3mm]
& & &- \displaystyle{\frac{\mi \Omega}{2 \sqrt{\Delta^2 + \Omega^2}}} \sin \left( \sqrt{\Delta^2 + \Omega^2}t \right) \me^{\mi \omega_{\textrm{L}} t}\\[3mm]
&\langle \hat{A}_{01} \left( t \right) \rangle &=& -\displaystyle{\frac{\Omega \Delta}{\Delta^2 + \Omega^2}}\sin^2 \left( \frac{1}{2} \sqrt{\Delta^2 + \Omega^2} t \right) \me^{-\mi \omega_{\textrm{L}} t}\\[3mm]
& & &+ \displaystyle{\frac{\mi \Omega}{2 \sqrt{\Delta^2 + \Omega^2}}} \sin \left( \sqrt{\Delta^2 + \Omega^2}t \right) \me^{-\mi \omega_{\textrm{L}} t}.
\label{eq:Occupation Probabilities}
\end{array}
\end{align}
The diagonal components of the solution can be interpreted as occupation probabilities $\langle \hat{A}{}_{00} \left( t \right) \rangle = p_0 \left( t \right)$ and $\langle \hat{A}{}_{11} \left( t \right) \rangle = p_1 \left( t \right)$ and the nondiagonal elements are needed to compute the dipole moment, which is given in Eq.~\eqref{eq:Dipole Moment General} by
\begin{align}
\begin{array}{llll}
&\langle \hat{\vec{d}} \left( t \right) \rangle &=& \left[ \langle \hat{A}_{10} \left( t \right) \rangle + \langle \hat{A}_{01} \left( t \right) \rangle \right] \vec{d}\\[2mm]
& &=& - \displaystyle{\frac{2 \Omega \Delta}{\Delta^2 + \Omega^2}} \sin^2 \left( \frac{1}{2} \sqrt{\Delta^2 + \Omega^2} t \right) \cos \left( \omega_{\textrm{L}} t \right) \vec{d}\\[2mm]
& & &+ \displaystyle{\frac{\Omega}{\sqrt{\Delta^2 + \Omega^2}}} \sin \left( \sqrt{\Delta^2 + \Omega^2} t \right) \sin \left( \omega_{\textrm{L}} t \right) \vec{d}.
\label{eq:Dipole Moment Bloch Equation}
\end{array}
\end{align}
The second part of the dipole moment \eqref{eq:Dipole Moment Bloch Equation} dominates over the first part in case of resonance $\Delta = 0$, whereas the first part plays an important role for a large detuning $\Delta \gg \Omega$.\\
The occupation probabilities $\langle \hat{A}_{11} \left( t \right) \rangle$ and $\langle \hat{A}_{00} \left( t \right) \rangle$ and the off-diagonal solutions $\langle \hat{A}_{10} \left( t \right) \rangle$ \eqref{eq:Occupation Probabilities} feature the dressed frequency $\sqrt{\Delta^2 + \Omega^2}$. The off-diagonal solutions oscillate with the laser frequency $\omega_{\textrm{L}}$, which will be seen to govern the oscillation frequency of the laser-induced Casimir--Polder potential. Ref.~\cite{Hughes:2013} gives a detailed analysis of the driving frequencies revealing the appearance of Mollow triplets \cite{Mollow:1969} consisting of the three frequencies $\omega_{\textrm{L}}$, $\omega_{\textrm{L}} + \Omega$ and $\omega_{\textrm{L}} - \Omega$, which are shifted by the Rabi frequency $\Omega$. Since in our case $\omega_{\textrm{L}} \gg \Omega$, we neglect the effect stemming from the Mollow triplet in the following analysis.

\subsubsection{Force of the Free Laser Field}
The potential of the free laser field is given in Eq.~\eqref{eq:Force Free Electric Laser Field} and has to be evaluated for the dipole moment \eqref{eq:Dipole Moment Bloch Equation}. After averaging over fast oscillating terms with the laser frequency $\omega_{\textrm{L}}$, we obtain the free electric force
\begin{multline}
U^{\textrm{el}}_{\textrm{L}} \left( \vec{r}{}_{\textrm{A}}, t \right) = \frac{1}{2} \vec{E} \left( \vec{r}{}_{\textrm{A}} \right) \sprod \vec{d} \frac{\Delta \Omega}{\Delta^2 + \Omega^2} \sin^2 \left( \frac{1}{2} \sqrt{\Delta^2 + \Omega^2} t \right).
\end{multline}
This result for the potential can be compared to the respective perturbative result \eqref{eq:Free Electric Force Perturbation} for a large detuning $\Delta \gg \Omega$. By applying the definition of the Rabi frequency $\Omega$ and the detuning $\Delta$ \eqref{eq:Definition Frequency} and averaging the expression $\sin^2 \left( \frac{1}{2} \sqrt{\Delta^2 + \Omega^2} t \right)$ to $\frac{1}{2}$ we obtain
\begin{equation}
U^{\textrm{el}}_{\textrm{L}} \left( \vec{r}{}_{\textrm{A}}, t \right) \approx \frac{1}{12} \frac{d^2 \vec{E}^2 \left( \vec{r}_{\textrm{A}} \right)}{\hbar \Delta},
\end{equation}
where we again assumed an isotropic atomic state. This result agrees with the respective result from the perturbative approach \eqref{eq:Free Electric Force Perturbation}.

\subsubsection{Casimir--Polder Potential}
To compute the laser-driven Casimir--Polder potential
\begin{multline}
U^{\textrm{BE}}_{\textrm{LCP}} =\\
- \frac{\mi \mu_0}{2 \pi} \int\limits^{\infty}_0 \dif \omega \omega^2 \int\limits^t_0 \dif \tau \me^{- \mi \omega \left( t-\tau \right)} \langle \hat{\vec{d}} \left( t \right) \sprod \mathrm{Im} \tens{G} \left( \vec{r}_{\textrm{A}}, \vec{r}_\textrm{A}, \omega \right) \sprod \hat{\vec{d}} \left( \tau \right) \rangle\\
+\frac{\mi \mu_0}{2 \pi} \int\limits^{\infty}_0 \dif \omega \omega^2 \int\limits^t_0 \dif \tau \me^{\mi \omega \left( t-\tau \right)} \langle \hat{\vec{d}} \left( \tau \right) \sprod \mathrm{Im} \tens{G} \left( \vec{r}_{\textrm{A}}, \vec{r}_\textrm{A}, \omega \right) \sprod \hat{\vec{d}} \left( t \right) \rangle
\end{multline}
one needs the correlation functions of the atomic flip operators $\langle \hat{A}_{10} \left( t \right) \hat{A}_{01} \left( \tau\right) \rangle$ and $\langle \hat{A}_{01} \left( t \right) \hat{A}_{10} \left( \tau\right) \rangle$ evaluated at time $t$ and $\tau$. Neglecting fast oscillating terms for $t \approx \tau$ cancels the correlation functions $\langle \hat{A}_{10} \left( t \right) \hat{A}_{10} \left( \tau\right) \rangle$ and $\langle \hat{A}_{01} \left( t \right) \hat{A}_{01} \left( \tau\right) \rangle$ and by making use of the relation \eqref{eq:Delta Function}, one obtains under the initial conditions from Sec.~\ref{sec:Dynamics and Dipole Moments}
\begin{align}
\begin{array}{lll}
&\langle \hat{A}_{10} \left( t \right) \hat{A}_{01} \left( \tau \right) \rangle &= \me^{\mi \omega_{\textrm{L}} \left( t-\tau \right)} \displaystyle{\frac{\Omega^2}{\Delta^2 + \Omega^2}} \sin^2 \left( \frac{1}{2} \sqrt{\Delta^2 + \Omega^2} t \right)\\[3mm]
&\langle \hat{A}_{01} \left( t \right) \hat{A}_{10} \left( \tau \right) \rangle &= \me^{-\mi \omega_{\textrm{L}} \left( t-\tau \right)} \left[ \displaystyle{\frac{\Delta^2}{\Delta^2 + \Omega^2}} \right.\\[3mm]
& &\left. + \displaystyle{\frac{\Omega^2}{\Delta^2 + \Omega^2}} \cos^2 \left( \frac{1}{2} \sqrt{\Delta^2 + \Omega^2} t \right) \right].
\label{eq:Probabilities Atomic Flip Operator}
\end{array}
\end{align}
In the RWA picture these correlation functions are named $\langle \hat{\tilde{A}}{}_{10} \left( t \right) \hat{\tilde{A}}{}_{01} \left( \tau \right) \rangle$ and $\langle \hat{\tilde{A}}{}_{01} \left( t \right) \hat{\tilde{A}}{}_{10} \left( \tau \right) \rangle$. These correlation functions are identical to the occupation probabilities \eqref{eq:Occupation Probabilities} of the system: $\langle \hat{A}{}_{00} \left( t \right) \rangle = p_0 \left( t \right) = \langle \hat{\tilde{A}}{}_{01} \left( t \right) \hat{\tilde{A}}{}_{10} \left( \tau \right) \rangle$ and $\langle \hat{A}{}_{11} \left( t \right) \rangle = p_1 \left( t \right) = \langle \hat{\tilde{A}}{}_{10} \left( t \right) \hat{\tilde{A}}{}_{01} \left( \tau \right) \rangle$. Therefore the total potential can be written in terms of occupation probabilities. The total potential in terms of the occupation probabilities reads
\begin{equation}
U^{\textrm{BE}}_{\textrm{LCP}} = p_0 \left( t \right) U_0 + p_1 \left( t \right) U_1
\label{eq:Driven Casimir-Polder Potential Bloch Equation}
\end{equation}
with the potentials for the ground state $U_0$ and the excited state $U_1$ given by
\begin{align}
\begin{array}{llll}
&U_0 &=& \displaystyle{\frac{\mu_0}{\pi}} \int\limits^{\infty}_0{\dif \xi \frac{\omega_{\textrm{L}} \xi^2}{\xi^2 + \omega^2_{\textrm{L}}} \vec{d} \sprod \tens{G} \left( \vec{r}{}_{\textrm{A}}, \vec{r}{}_{\textrm{A}}, \mi \xi \right) \sprod \vec{d}}\\[5mm]
&U_1 &=& -\displaystyle{\frac{\mu_0}{\pi}} \int\limits^{\infty}_0{\dif \xi \frac{\omega_{\textrm{L}} \xi^2}{\xi^2 + \omega^2_{\textrm{L}}} \vec{d} \sprod \tens{G} \left( \vec{r}{}_{\textrm{A}}, \vec{r}{}_{\textrm{A}}, \mi \xi \right) \sprod \vec{d}}\\[5mm]
& & &- \mu_0 \omega^2_{\textrm{L}} \vec{d} \sprod \mathrm{Re} \tens{G} \left( \vec{r}{}_{\textrm{A}}, \vec{r}{}_{\textrm{A}}, \omega_{\textrm{L}} \right) \sprod \vec{d},
\label{eq:Driven Casimir-Polder Potential Bloch Equation Contributions}
\end{array}
\end{align}
where we have used the identity $\langle \hat{\tilde{A}}{}_{10} \left( t \right) \hat{\tilde{A}}{}_{01} \left( \tau \right) \rangle = \langle \hat{\tilde{A}}{}_{10} \left( \tau \right) \hat{\tilde{A}}{}_{01} \left( t \right) \rangle$. The final expression consists of a nonresonant contribution under the integral and a resonant one.\\
In the large-detuning limit $\Delta \gg \Omega$, where the atomic polarizability \eqref{eq:Atomic Polarizability Detuning} is applied, and after a time-average $\sin^2 \left( \omega_{\textrm{L}} t \right) \rightarrow \frac{1}{2}$ the resonant contribution of this expression agrees exactly with Eq.~\eqref{eq:Driven Casimir-Polder Potential Perturbation}.\\
If we set the electric driving field $\vec{E} \left( \vec{r}_{\textrm{A}} \right)$ equal to $0$ the probabilities reduce to $p_0 \left( t \right) = 1$ and $p_1 \left( t \right) = 0$ and Eq.~\eqref{eq:Driven Casimir-Polder Potential Bloch Equation} with Eq.~\eqref{eq:Driven Casimir-Polder Potential Bloch Equation Contributions} is equal to the standard ground state Casimir--Polder potential, given the laser frequency is set to the atomic transition frequency $\omega_{\textrm{L}} = \tilde{\omega}_{10}$ \cite{Buhmann_Book_1, Buhmann_Book_2}. The potential shows only a nonresonant integral term.\\
If the atom is initially in its excited state $\langle \hat{A}_{11} \left( 0 \right) \rangle = 1$, $\langle \hat{A}_{11} \left( 0 \right) \rangle = 0$ and we set the electrical driving field to 0, the probabilities are $p_1 \left( t \right) = 1$ and $p_0 \left( t \right) = 0$, the potential \eqref{eq:Driven Casimir-Polder Potential Bloch Equation} with Eq.~\eqref{eq:Driven Casimir-Polder Potential Bloch Equation Contributions} is identical to the Casimir--Polder potential for an atom in its excited state. In this case the potential is composed of a resonant part containing the transition frequency and a nonresonant contribution.

\section{Atom Near a Plane Surface}
\label{sec:Atom Near a Plane Surface}
\begin{figure}[!ht]
\centerline{\includegraphics[width=\columnwidth]{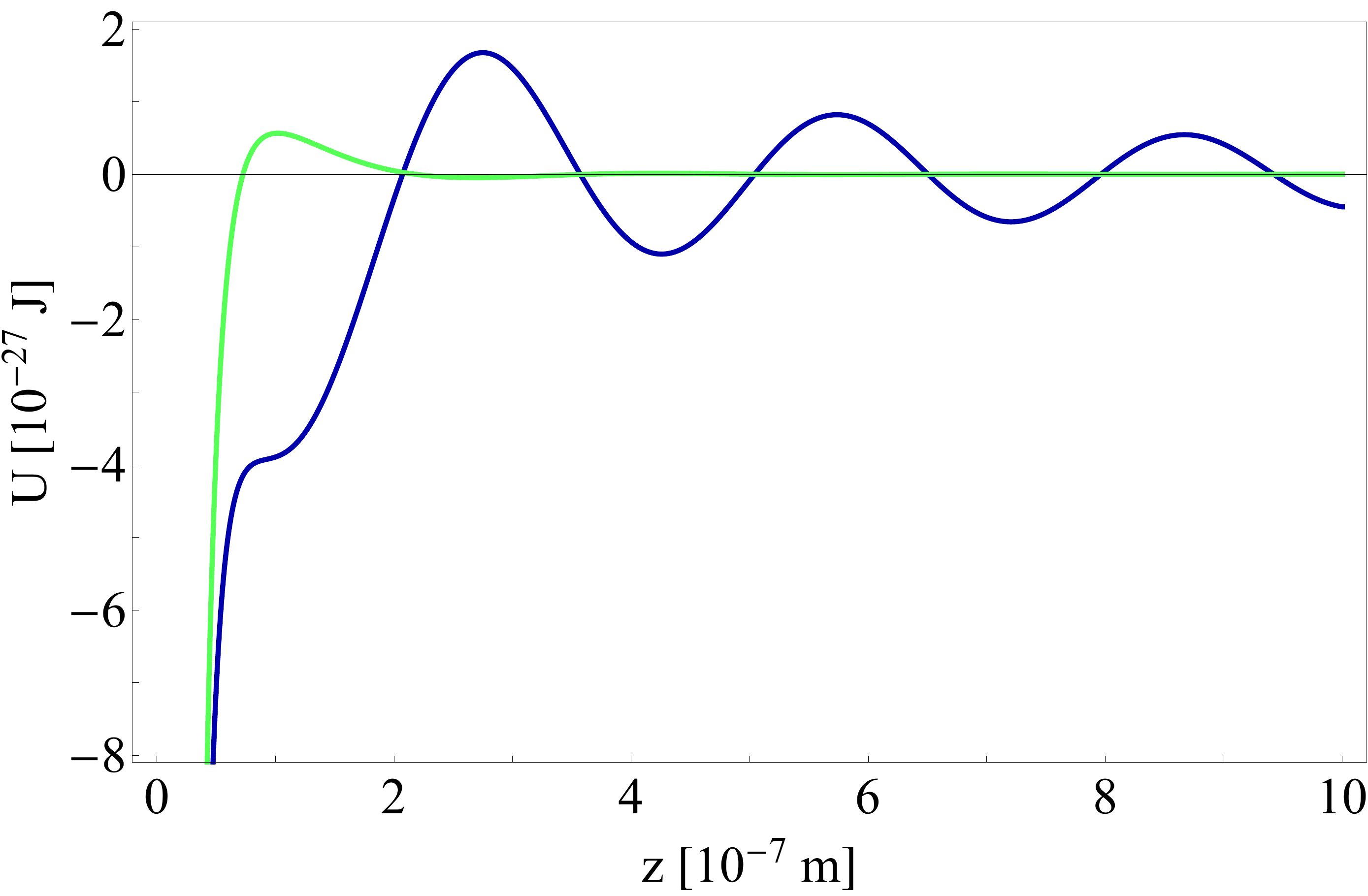}}
\caption{Total Casimir--Polder potential from the perturbative approach for a Na-atom close to a surface driven by a laser with an intensity of $I = 5 \: \textrm{W}/\textrm{cm}^2$ \eqref{eq:Result Casimir-Polder Potential Perturbation} ($\textcolor{Blue}{\hdashrule[0.5ex][x]{0.6cm}{1pt}{}}$) and the potential with only the contribution proportional to $z^{-3}$ \eqref{eq:Result Casimir-Polder Potential Perturbation Perreault} ($\textcolor{LimeGreen}{\hdashrule[0.5ex][x]{0.6cm}{1pt}{}}$).}
\label{fig:Figure1}
\end{figure}

\begin{figure}[!ht]
\centerline{\includegraphics[width=\columnwidth]{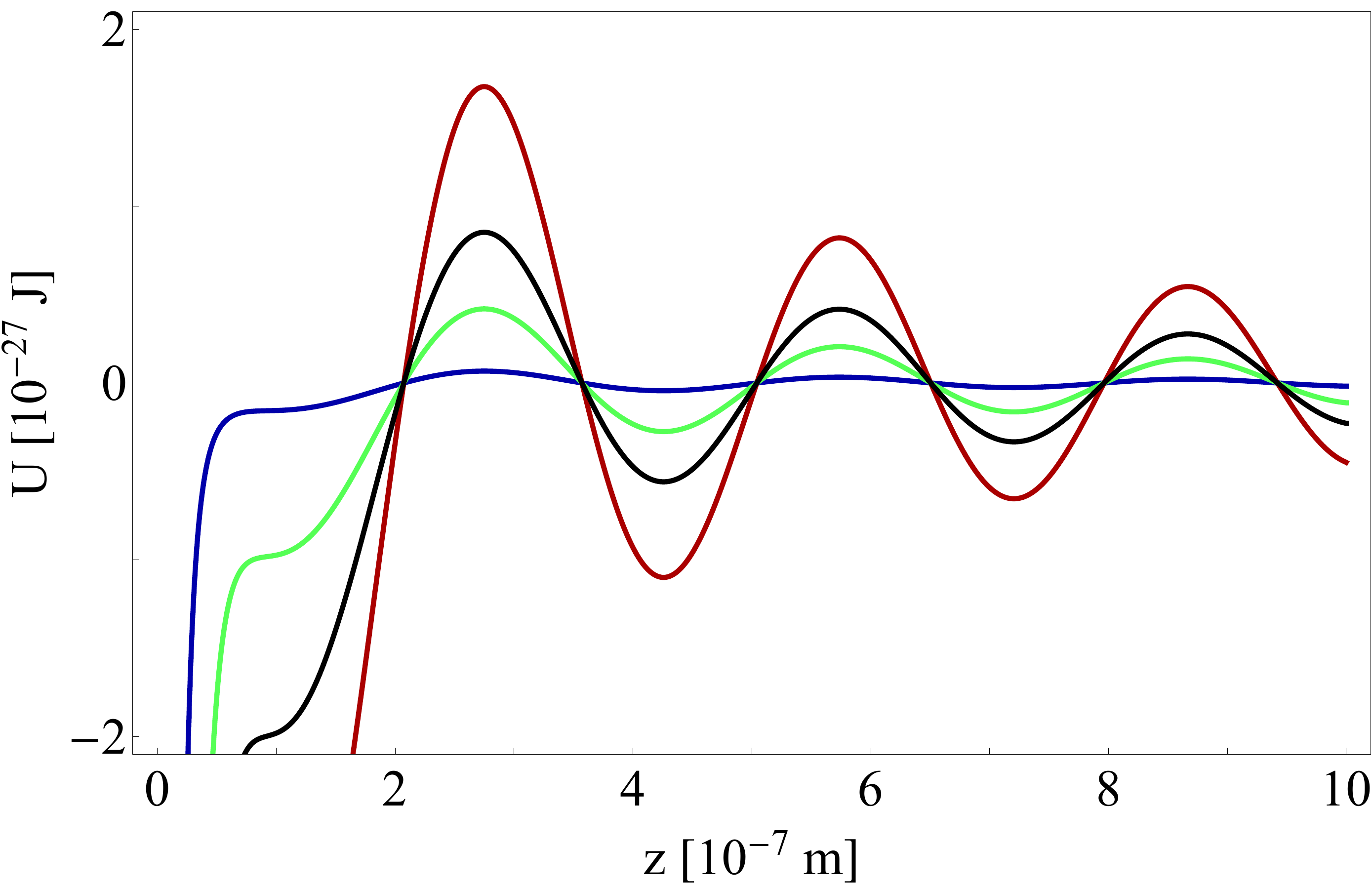}}
\caption{Total Casimir--Polder potential from the perturbative approach for a Na-atom close to a surface driven by a laser with an intensity of $I = 5 \: \textrm{W}/\textrm{cm}^2$ \eqref{eq:Result Casimir-Polder Potential Perturbation} with the detunings $5 \left( \omega_{\textrm{L}} - \tilde{\omega}_{10} \right)$ ($\textcolor{Blue}{\hdashrule[0.5ex][x]{0.6cm}{1pt}{}}$), $2 \left( \omega_{\textrm{L}} - \tilde{\omega}_{10} \right)$ ($\textcolor{LimeGreen}{\hdashrule[0.5ex][x]{0.6cm}{1pt}{}}$) and $\omega_{\textrm{L}} - \tilde{\omega}_{10}$ ($\textcolor{Red}{\hdashrule[0.5ex][x]{0.6cm}{1pt}{}}$) and the undriven Casimir--Polter \eqref{eq:Casimir-Polder Potential Undriven} ($\textcolor{Black}{\hdashrule[0.5ex][x]{0.6cm}{1pt}{}}$).}
\label{fig:Figure2}
\end{figure}

\begin{figure}[!ht]
\centerline{\includegraphics[width=\columnwidth]{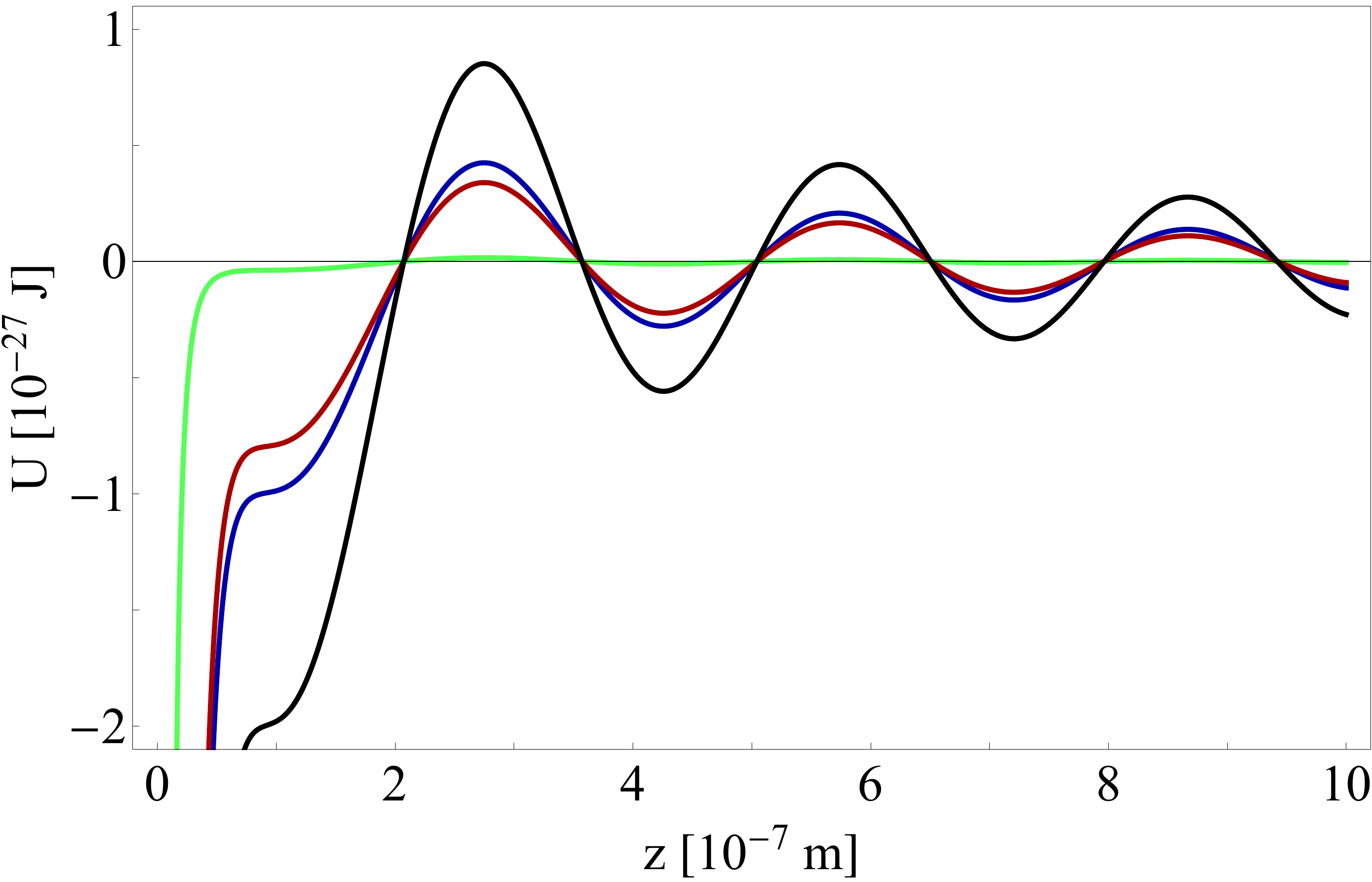}}
\caption{Total Casimir--Polder potential from the Bloch equation approach for a Na-atom close to a surface driven by a laser with an intensity of $I = 5 \: \textrm{W}/\textrm{cm}^2$ \eqref{eq:Result Casimir-Polder Potential Bloch Equation} with the detunings $0.1 \left( \omega_{\textrm{L}} - \tilde{\omega}_{10} \right)$ ($\textcolor{Blue}{\hdashrule[0.5ex][x]{0.6cm}{1pt}{}}$), $10 \left( \omega_{\textrm{L}} - \tilde{\omega}_{10} \right)$ ($\textcolor{LimeGreen}{\hdashrule[0.5ex][x]{0.6cm}{1pt}{}}$) and $\omega_{\textrm{L}} - \tilde{\omega}_{10}$ ($\textcolor{Red}{\hdashrule[0.5ex][x]{0.6cm}{1pt}{}}$) and the undriven Casimir--Polter \eqref{eq:Casimir-Polder Potential Undriven} ($\textcolor{Black}{\hdashrule[0.5ex][x]{0.6cm}{1pt}{}}$).}
\label{fig:Figure3}
\end{figure}

\begin{figure}[!ht]
\centerline{\includegraphics[width=\columnwidth]{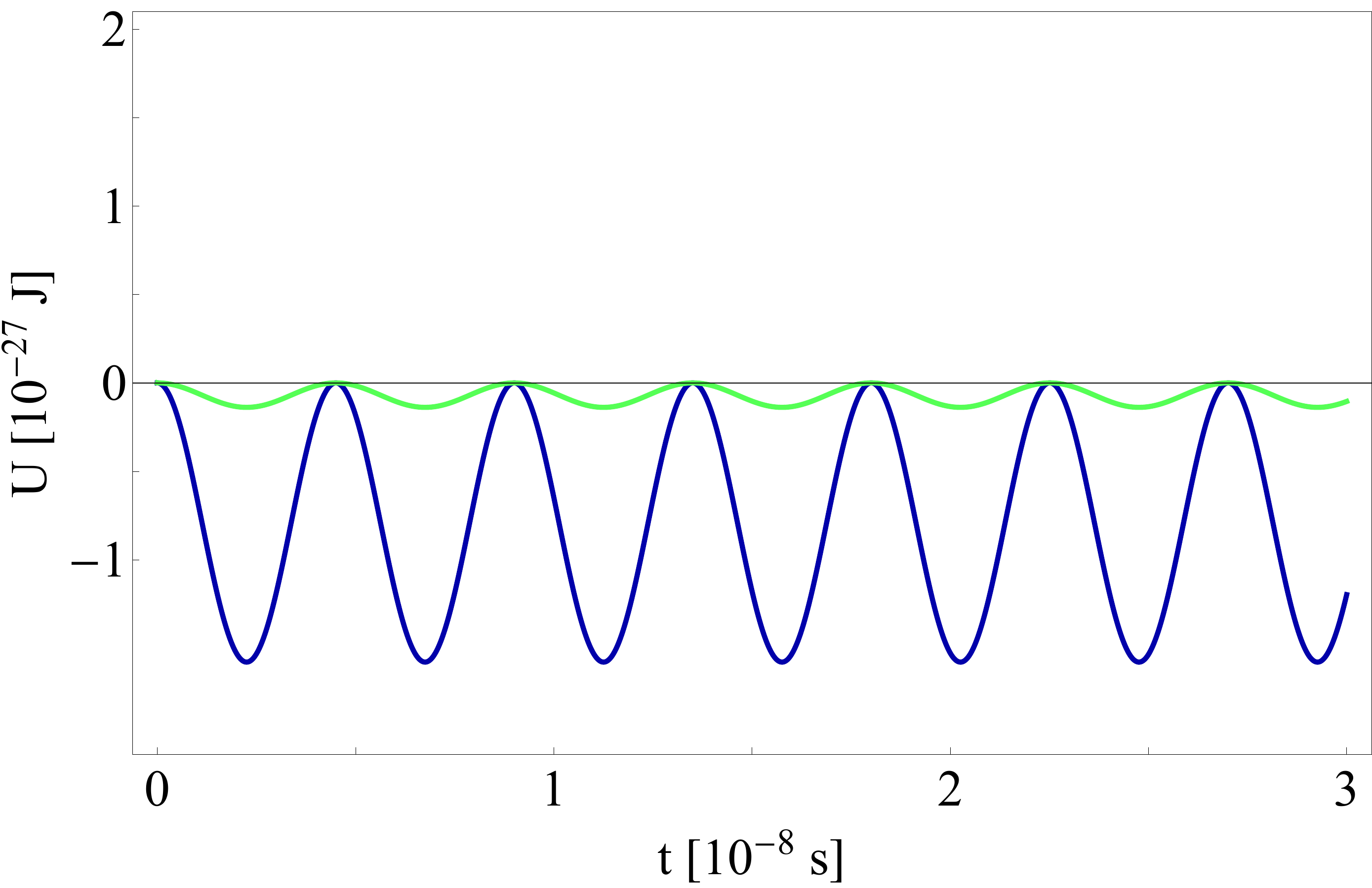}}
\caption{Time-dependent Casimir--Polder potential from the Bloch equation approach for a Na-atom close to a surface driven by a laser with an intensity of $I = 5 \: \textrm{W}/\textrm{cm}^2$ \eqref{eq:Result Casimir-Polder Potential Bloch Equation} for a distance of $z = 2 \times 10^{-7} \: \textrm{m}$ ($\textcolor{LimeGreen}{\hdashrule[0.5ex][x]{0.6cm}{1pt}{}}$) and a distance of $z = 10^{-7} \: \textrm{m}$ ($\textcolor{Blue}{\hdashrule[0.5ex][x]{0.6cm}{1pt}{}}$).}
\label{fig:Figure4}
\end{figure}

\begin{figure}[!ht]
\centerline{\includegraphics[width=\columnwidth]{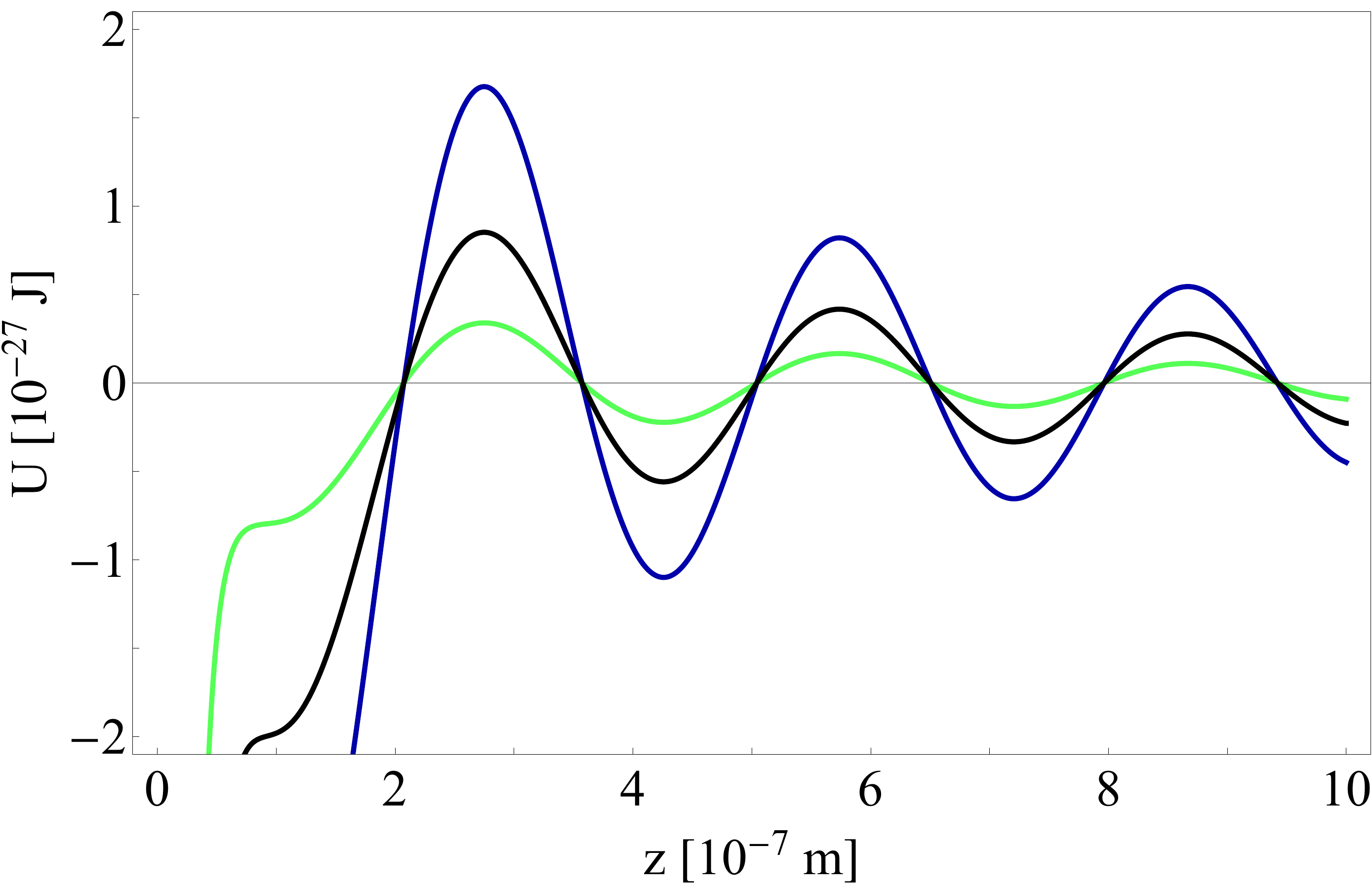}}
\caption{Total Casimir--Polder potential for a Na-atom close to a surface driven by a laser with an intensity of $I = 5 \: \textrm{W}/\textrm{cm}^2$ for the perturbative approach \eqref{eq:Result Casimir-Polder Potential Perturbation} ($\textcolor{Blue}{\hdashrule[0.5ex][x]{0.6cm}{1pt}{}}$) and the Bloch equation approach \eqref{eq:Result Casimir-Polder Potential Bloch Equation} ($\textcolor{LimeGreen}{\hdashrule[0.5ex][x]{0.6cm}{1pt}{}}$). This is compared to the Casimir--Polder potential for an undriven atom \eqref{eq:Casimir-Polder Potential Undriven} ($\textcolor{Black}{\hdashrule[0.5ex][x]{0.6cm}{1pt}{}}$).}
\label{fig:Figure5}
\end{figure}

We want to evaluate the driven Casimir--Polder potential for the atom \eqref{eq:Driven Casimir-Polder Potential Perturbation} under the influence of the driving laser field \eqref{eq:Electric Driving Field} for a specific choice of applied electric field and geometry. In Ref.~\cite{Fuchs:2018} we apply the result for the laser-induced Casimir--Polder potential in Eq.~\eqref{eq:Driven Casimir-Polder Potential Perturbation} to a specific laser driving field, namely an evanescent laser beam under realistic experimental conditions and compare its contribution to the sum of the light-potential and the Casimir--Polder potential. Ref.~\cite{Perreault:2008} studies this setup for the electrical driving field
\begin{equation}
\vec{E} \left( \vec{r}_{\textrm{A}} \right) = E_0 \left( \vec{r}_{\textrm{A}} \right) \begin{pmatrix} \sin \left( \theta \right) \\ 0 \\ \cos \left( \theta \right) \end{pmatrix}.
\label{eq:Dipole Moments Angles}
\end{equation}
The angle $\theta$ is between the z-axis and the orientation of the field $\vec{E} \left( \vec{r}_{\textrm{A}} \right)$. The unpolarized dipole moment induced by this field is aligned in the same direction and its image dipole differs by a sign in the x-component. We study the setup for a perfectly conducting mirror with the reflective coefficients $r_{\textrm{s}} = -1$ and $r_{\textrm{p}} = 1$ leading to the components of the scattering part of the Green's tensor 
\begin{multline}
\tens{G}^{(1)}_{xx} \left( \vec{r}, \vec{r}, \omega \right) = \tens{G}^{(1)}_{yy} \left( \vec{r}, \vec{r}, \omega \right)\\[2mm]
=\frac{\omega}{32 \pi c} \left[ \left( \frac{c}{\omega z} \right)^3 - 2 \mi \left( \frac{c}{\omega z} \right)^2 - 4 \left( \frac{c}{\omega z} \right) \right] \me^{ \frac{2 \mi \omega z}{c}}\\[2mm]
\tens{G}^{(1)}_{zz} \left( \vec{r}, \vec{r}, \omega \right) = \frac{\omega}{16 \pi c} \left[ \left( \frac{c}{\omega z} \right)^3 - 2 \mi \left( \frac{c}{\omega z} \right)^2 \right] \me^{\frac{2 \mi \omega z}{c}}.
\label{eq:Greens Tensor Scattering Part}
\end{multline}
The non-diagonal elements of the Green's tensor are equal to $0$. This result reflects the interaction of the dipole moment \eqref{eq:Dipole Moments Angles} with itself mediated by the presence of the surface of the perfectly conducting mirror with Green's tensor \eqref{eq:Greens Tensor Scattering Part}. 
By making use of Eq.~\eqref{eq:Driven Casimir-Polder Potential Perturbation} the Casimir--Polder potential for the laser-driven atom eventually reads
\begin{multline}
U^{\textrm{per}}_{\textrm{LCP}} \left( \vec{r}_{\textrm{A}} \right) = - \displaystyle{\frac{\mu_0 \omega^3_{\textrm{L}} \alpha^2_n \left( \omega_{\textrm{L}} \right) E^2_0 \left( \vec{r}_{\textrm{A}} \right)}{64 \pi c}}\\
\left\{ \left[ 1 + \cos^2 \left( \theta \right) \right] \left( \frac{c}{\omega_{\textrm{L}} z} \right)^3 \cos \left( {\frac{2 \omega_{\textrm{L}} z}{c}} \right) \right.\\
\left. + 2 \left[ 1 + \cos^2 \left( \theta \right) \right] \left( \frac{c}{\omega_{\textrm{L}} z} \right)^2 \sin \left( {\frac{2 \omega_{\textrm{L}} z}{c}} \right) \right.\\
\left. - 4 \sin^2 \left( \theta \right) \left( \frac{c}{\omega_{\textrm{L}} z} \right) \cos \left( {\frac{2 \omega_{\textrm{L}} z}{c}} \right) \right\}.
\label{eq:Result Casimir-Polder Potential Perturbation}
\end{multline}
We have again used real and isotropic atomic polarizabilities. The Casimir-Polder potential for the laser-induced electric field can be approximated in the retarded limit $\left( \omega_{\textrm{L}} z/c \gg 1 \right)$ and in the nonretarded limit $\left( \omega_{\textrm{L}} z/c \ll 1 \right)$
\begin{align}
\begin{array}{lll}
&U^{\textrm{per}}_{\textrm{LCP}} \left( \vec{r}_{\textrm{A}} \right) &= \begin{cases} \displaystyle{\frac{\mu_0 \omega_{\textrm{L}}^2 \alpha^2_n \left( \omega_{\textrm{L}} \right) E^2_0 \left( \vec{r}_{\textrm{A}} \right)}{16 \pi z} \sin^2 \left( \theta \right) \cos \left( \frac{2 \omega_{\textrm{L}} z}{c} \right)},\\
\displaystyle{\frac{\omega_{\textrm{L}} z}{c} \gg 1}, \\[2mm]
-\displaystyle{\frac{\mu_0 \omega_{\textrm{L}} \alpha^2_n \left( \omega_{\textrm{L}} \right) E^2_0 \left( \vec{r}_{\textrm{A}} \right) c^2}{64 \pi z^3}} \left[ 1 + \cos^2 \left( \theta \right) \right],\\
\displaystyle{\frac{\omega_{\textrm{L}} z}{c} \ll 1}. \end{cases}
\label{eq:Result Casimir-Polder Potential Perturbation Limits}
\end{array}
\end{align}
We compare the expression for the laser-induced Casimir--Polder potential in Eq.~\eqref{eq:Result Casimir-Polder Potential Perturbation} using macroscopic QED with the results from Ref.~\cite{Perreault:2008}, wherein both the electric field and the induced atom-surface interaction potential are computed using the image dipole method.\\
The monochromatic external field \eqref{eq:Electric Driving Field} acts on the atom, whose dipole moment is then aligned in the same direction. The induced electric field of the atom \eqref{eq:Electric Field Induced} is given by
\begin{multline}
\vec{E} \left( \vec{r}_{\textrm{A}}, t \right) = \frac{1}{2} \left[ 3 \cos \left( \theta \right) \vec{e}_z - \vec{e}_p \right] \frac{\alpha_n \left( \omega_{\textrm{L}} \right) E_0 \left( \vec{r}{}_{\textrm{A}} \right)}{4 \pi \epsilon_0 \left( 2z \right)^3}\\
\left[ \me^{-\mi \omega_{\textrm{L}} t} \me^{\frac{2 \mi \omega_{\textrm{L}} z}{c}} + \me^{\mi \omega_{\textrm{L}} t} \me^{- \frac{2 \mi \omega_{\textrm{L}} z}{c}} \right]
\end{multline}
containing the time-dependency of the electric driving field \eqref{eq:Electric Driving Field}. By using the unitary vectors in the z-direction $\vec{e}_z$ and the direction of the image dipole moment $\vec{e}_p$ our result is identical to the respective expression in Ref.~\cite{Perreault:2008}. The respective atom-surface interaction potential in this notation using Eq.~\eqref{eq:Result Casimir-Polder Potential Perturbation} is given by
\begin{equation}
U_{\textrm{LCP}} \left( \vec{r}_{\textrm{A}} \right) = -\frac{ \alpha^2 \left( \omega_{\textrm{L}} \right) E^2_0 \left( \vec{r}_{\textrm{A}} \right)}{64 \pi \epsilon_0 z^3} \left[ 1 + \cos^2 \left( \theta \right) \right] \cos \left( \frac{2 \omega_{\textrm{L}} z}{c} \right)
\label{eq:Result Casimir-Polder Potential Perturbation Perreault}
\end{equation}
and has lost the time-dependent terms. Equation \eqref{eq:Result Casimir-Polder Potential Perturbation Perreault} agrees perfectly with the respective result from Ref.~\cite{Perreault:2008}.\\
In Ref.~\cite{Perreault:2008} it is stated that the terms in $z^{-2}$ and $z^{-1}$ are neglected in the near-field regime. This expression is identified with the nonretarded limit of the laser-driven Casimir--Polder potential in Eq.~\eqref{eq:Result Casimir-Polder Potential Perturbation Limits}, which is proportional to $z^{-3}$. Figure \ref{fig:Figure1} shows that Eq.~\eqref{eq:Result Casimir-Polder Potential Perturbation Perreault} is not sufficient for the interaction potential \eqref{eq:Result Casimir-Polder Potential Perturbation}.\\
The obtained equations are evaluated by making use of the example presented in Ref.~\cite{Perreault:2008}. We have partly used more accurate values stemming from an increased precision of measurements. The static atomic polarizability is given by the expression $\alpha_{\textrm{DC}} = e^2/m \tilde{\omega}{}^2_{10}$ with the electron mass $m$ and the resonance frequency $\tilde{\omega}{}_{10}$ and a value of $\alpha_{\textrm{DC}}/(4 \pi \epsilon_0) = 24 \vprod 10^{-30} \: \textrm{m}^3$ is delivered. We used a more recent experimentally determined value of $\alpha_{\textrm{DC}}/(4 \pi \epsilon_0) = 24.11 \vprod 10^{-30} \: \textrm{m}^3$ \cite{Holmgren:2010} for our calculations. The laser intensity is $I=\epsilon_0 c |E_0 \left( \vec{r}_{\textrm{A}} \right)|^2/2=5 \: \textrm{W}/\textrm{cm}^2$ and the detuning between laser frequency $\omega_{\textrm{L}}$ and the resonance frequency $\tilde{\omega}{}_{10}$ has a value of $\omega_{\textrm{L}} - \tilde{\omega}_{10} = 2 \pi \vprod 100 \: \textrm{MHz}$. This yields values for the atomic transition frequency of $\tilde{\omega}_{10} = 3.24 \vprod 10^{15} \: 1/\textrm{s}$ and the dipole moment $d=3.71 \vprod 10^{-29} \: \textrm{Cm}$. The detuning is seven orders of magnitude smaller than the atomic transition frequency $\tilde{\omega}_{10}$ and  $0.29$ of the value of the Rabi frequency $\Omega$ \eqref{eq:Definition Frequency}. We assume the dipole to be aligned along the x-axis $\theta = \pi/2$ and thus parallel to the surface.\\
Using these parameters the light-force potential \eqref{eq:Force Free Electric Laser Field} has a value of $U_L = -1.30 \vprod 10^{-27} \: \textrm{J}$, which is attractive and in the range of the Casimir--Polder potential.\\
Figure \ref{fig:Figure1} compares the total expression of the driven Casimir--Polder potential \eqref{eq:Result Casimir-Polder Potential Perturbation} with Eq.~\eqref{eq:Result Casimir-Polder Potential Perturbation Perreault}, which is identical with the nonretarded limit of Eq.~\eqref{eq:Result Casimir-Polder Potential Perturbation}. Consequently, we see good agreement between both curves at small distances. Nevertheless, the magnitude of this approximation from Ref.~\cite{Perreault:2008} does not agree well with the result obtained from Eq.~\eqref{eq:Result Casimir-Polder Potential Perturbation}.
In a next step, Eq.~\eqref{eq:Result Casimir-Polder Potential Perturbation} is evaluated for several detuning values and is compared with the Casimir--Polder potential of the undriven atom in its excited state
\begin{multline}
U_{\textrm{CP}} \left( \vec{r}_{\textrm{A}} \right) = - \mu_0 \tilde{\omega}^2_{10} \vec{d} \cdot \textrm{Re} \tens{G}^{(1)} \left( \vec{r}_{\textrm{A}}, \vec{r}_{\textrm{A}}, \tilde{\omega}_{10} \right) \cdot \vec{d}\\[3mm]
= - \frac{\mu_0 \tilde{\omega}^3_{10} d^2}{96 \pi c} \left\{ \left( \frac{c}{\tilde{\omega}_{10} z} \right)^3 \cos \left( {\frac{2 \tilde{\omega}_{10} z}{c}} \right) \right.\\
\left. + 2 \left( \frac{c}{\tilde{\omega}_{10} z} \right)^2 \sin \left( {\frac{2 \tilde{\omega}_{10} z}{c}} \right) \right.\\[3mm]
\left. - 4 \left( \frac{c}{\tilde{\omega}_{10} z} \right) \cos \left( {\frac{2 \tilde{\omega}_{10} z}{c}} \right) \right\}.
\label{eq:Casimir-Polder Potential Undriven}
\end{multline}
The dipole moment is chosen to be aligned along the x-axis with $d^2_x = \frac{1}{3} d^2$ to establish the same conditions as for the laser-driven potential. The Casimir--Polder potential for the undriven atom in its excited state can also be approximated in its retarded $\left( \tilde{\omega}_{10} z/c \gg 1 \right)$ and nonretarded limit $\left( \tilde{\omega}_{10} z/c \ll 1 \right)$
\begin{align}
\begin{array}{lll}
&U_{\textrm{CP}} \left( \vec{r}_{\textrm{A}} \right) &= \begin{cases} \displaystyle{\frac{\mu_0 \tilde{\omega}^2_{10} d^2}{24 \pi z} \cos \left( \frac{2 \tilde{\omega}_{10} z}{c} \right)}, & \displaystyle{\frac{\tilde{\omega}_{10} z}{c} \gg 1}, \\[2mm] -\displaystyle{\frac{\mu_0 d^2 c^2}{96 \pi z^3}}, & \displaystyle{\frac{\tilde{\omega}_{10} z}{c} \ll 1}. \end{cases}
\end{array}
\end{align}
Since the perturbative approach assumes the atom to stay in its initial state during the atomic dynamics, the detuning must not be too small. Fig.~\ref{fig:Figure2} compares the driven Casimir--Polder potential from the perturbative approach, Eq.~\eqref{eq:Result Casimir-Polder Potential Perturbation}, with three different detuning values with the standard Casimir--Polder potential. There is good agreement between the driven potential with a detuning of $\omega_{\textrm{L}} - \tilde{\omega}_{10}$ and the standard Casimir--Polder potential. Since the detuning between the laser frequency $\omega_{\textrm{L}}$ and the atomic transition frequency $\tilde{\omega}_{10}$ is very small, all of the potentials are in phase.\\
The result for the driven Casimir--Polder potential following the approach using Bloch equations based on Eq.~\eqref{eq:Driven Casimir-Polder Potential Bloch Equation} from Sec.~\ref{sec:Bloch Equation} is evaluated using a dipole moment aligned with the electric field in Eq.~\eqref{eq:Dipole Moments Angles} and the scattering part of the Green's tensor \eqref{eq:Greens Tensor Scattering Part}. We obtain for the resonant contribution
\begin{multline}
U^{\textrm{BE}}_{\textrm{LCP}} \left( \vec{r}_A, t \right) = - \frac{\mu_0 \omega^3_{\textrm{L}} d^2}{96 \pi c} \frac{\Omega^2}{\Delta^2 + \Omega^2} \sin^2 \left( 2 \sqrt{\Delta^2 + \Omega^2} t \right)\\
\left\{ \left( \frac{c}{\omega_{\textrm{L}} z} \right)^3 \cos \left( \frac{2 \omega_{\textrm{L}} z}{c} \right) \right.\\
\left. + 2 \left( \frac{c}{\omega_{\textrm{L}} z} \right)^2 \sin \left( \frac{2 \omega_{\textrm{L}} z}{c} \right) \right.\\[3mm]
\left. - 4 \left( \frac{c}{\omega_{\textrm{L}} z} \right) \cos \left( \frac{2 \omega_{\textrm{L}} z}{c} \right) \right\}.
\label{eq:Result Casimir-Polder Potential Bloch Equation}
\end{multline}
We have assumed the dipole moment $\vec{d}$ to be isotropic with $d^2_x = d^2_y = d^2_z = \frac{1}{3} \vec{d}^2$. Since the laser-field strength is included in the Rabi frequency $\Omega$, the laser-driven Casimir--Polder potential \eqref{eq:Result Casimir-Polder Potential Bloch Equation} reaches a value of saturation for $\Omega \gg \Delta$, which is $\frac{1}{2}$ of the value of the standard undriven Casimir--Polder potential. This value represents an upper boundary to the increase of the potential due to an applied field.\\
In the retarded/nonretarded limit and after the averaging over time $U^{\textrm{BE}}_{\textrm{LCP}}$ approximates to
\begin{align}
\begin{array}{lll}
&U^{\textrm{BE}}_{\textrm{LCP}} \left( \vec{r}_{\textrm{A}} \right) &= \begin{cases} \displaystyle{\frac{\mu_0 \omega^2_{\textrm{L}} d^2}{48 \pi z} \frac{\Omega^2}{\Delta^2 + \Omega^2} \cos \left( \frac{2 \omega_{\textrm{L}} z}{c} \right)}, & \displaystyle{\frac{\omega_{\textrm{L}} z}{c} \gg 1}, \\[2mm] -\displaystyle{\frac{\mu_0 d^2 c^2}{192 \pi z^3}} \frac{\Omega^2}{\Delta^2 + \Omega^2}, & \displaystyle{\frac{\omega_{\textrm{L}} z}{c} \ll 1}. \end{cases}
\end{array}
\end{align}
Again we have assumed the dipole moment to be isotropic. The result \eqref{eq:Result Casimir-Polder Potential Bloch Equation} contains an additional time-dependency in contrast to the Eq.~\eqref{eq:Result Casimir-Polder Potential Perturbation}. By averaging over time, the distance-dependence of the potential can be investigated and compared to the off-resonant case. Fig.~\ref{fig:Figure3} compares the Casimir--Polder potential from the Bloch equation approach for several detunings \eqref{eq:Result Casimir-Polder Potential Bloch Equation} with the original Casimir--Polder potential \eqref{eq:Casimir-Polder Potential Undriven}.\\
The time-dependence of $U^{\textrm{BE}}_{\textrm{LCP}}$ can be studied by looking at different distances. In Fig.~\ref{fig:Figure4} we observe structures similar to Rabi oscillations.
The oscillations for two different distances of the atom from the surface have different amplitudes and different phases depending on the the sign and value of the potential at these distance values.\\
Fig.~\ref{fig:Figure5} compares the results for the driven Casimir--Polder potential from the perturbative approach, the Bloch equation approach and the undriven Casimir--Polder potential.
For the chosen detuning $\omega_{\textrm{L}} - \tilde{\omega}_{10}$, all curves are in phase and the result from the perturbative approach and the standard Casimir--Polder potential agree very well. The respective result from the Bloch equation approach is also in phase, but reaches not more than $1/2$ of the value of the undriven Casimir--Polder potential. For small distances all of the potentials are negative and will lead to an attractive force between the atoms and the surface. Whereas the perturbative approach is limited by the value of the detuning and would produce unphysical values in the opposite case, the Bloch equation method is valid for all detunings.

\section{Summary}
We have computed the Casimir--Polder potential of an atom in proximity of a surface that is driven by a monochromatic laser field. Applying a perturbative approach and using Bloch equations, we have compared the results with the standard Casimir--Polder potential caused by spontaneously arising polarizations and magnetizations.\\
Our calculations are formulated in the theory of macroscopic quantum electrodynamics (QED), that describes matter macroscopically by permittivity and permeability tensors. In Sec.~\ref{sec:Macroscopic Quantum Electrodynamics} we first established an expression for the electric field consisting of a free laser field and the field emitted by the atom close to the surface. The internal atomic dynamics is studied in the form of equations of motion for the atomic flip operator (Sec.~\ref{sec:Internal Atomic Dynamics}). From this point on, we have distinguished between a perturbative treatment, where the atom stays in its initial state during the dynamics (Secs.~\ref{sec:Perturbative Approach} and \ref{sec:Components of the Electric Potential}), and a different way using Bloch equations (Sec.~\ref{sec:Bloch Equation}). In the former case we computed the dipole moment of the laser-driven atom, found an expression for the electric field and used both expressions to obtain the respective potential. Since both the electric field and the dipole moment can be split into a free part connected to the field fluctuations and an laser-induced part, one can obtain the laser-driven and standard expression for the Casimir--Polder potentials. In the second approach we solved the Maxwell-Bloch equations and represented the laser-driven Casimir--Polder potential in terms of correlation functions. The final result agrees with the perturbative result in the large-detuning limit.\\
In Sec.~\ref{sec:Atom Near a Plane Surface} the results are applied to a dipole moment parallel to the surface induced by an electric laser field pointing in the same direction. The results are compared to the standard Casimir--Polder potential and agree very well with each other. The driven Casimir--Polder potential from the Bloch equation approach also agrees with the standard undriven potential, but can only reach a maximum of $1/2$ of its respective value in case of a small detuning. The perturbative approach is based on the assumption that the atom stays in its initial state during the atomic dynamics. Therefore this approach is restricted to large detunings, whereas the Bloch equation approach does not show such a restriction.\\
This derivation makes the artificial creation of the Casimir--Polder potential by using a driving laser-field possible. Nevertheless, it was shown that the laser-driven potential has an upper boundary which cannot be overcome. It is also to be expected that this effect will be increased by coupling several atoms in proximity of a surface to a laser field. This possible enhancement of Casimir--Polder potentials due to an applied electric field makes it seem possible to visualize particularly small effects being connected to Casimir--Polder potentials between a chiral object and a surface \cite{Barcellona:2016} or an atom and nonreciprocal material, such as a topological insulator \cite{Fuchs:2016}. Both of these materials have electromagnetic properties that are based on the coupling between electric and magnetic fields which are usually very small.\\
Ref.~\cite{Fuchs:2018} compares the laser-driven Casimir--Polder potential driven by an evanescent wave under experimentally realizable conditions with the usually assumed sum of the light potential and the standard Casimir--Polder potential and shows its significance thus proving the non-additivity of these two potentials.

\section{acknowledgments}
We acknowledge helpful discussions with Francesco Intravaia, Diego Dalvit and Ian Walmsley. This work was supported by the German Research Foundation (DFG, Grants BU 1803/3-1 and GRK 2079/1). S.Y.B is grateful for support by the Freiburg Institute of Advanced Studies.


%

\end{document}